\documentclass{aa}
\usepackage{graphicx} 
\usepackage{float}
\usepackage{txfonts}
\usepackage[breaklinks, colorlinks, allcolors=blue]{hyperref}
\usepackage{xcolor}
\usepackage{natbib}
\usepackage{epstopdf}
\usepackage{multirow,array}
\usepackage{caption}
\usepackage{longtable}
\usepackage{lscape}
\usepackage[normalem]{ulem}
\usepackage{amsmath,siunitx}
\usepackage{url}
\bibpunct{(}{)}{;}{a}{}{,}

\begin{document}

\title{STONKS first results: Long-term transients in the \textit{XMM-Newton} Galactic plane survey}
   \author{Robbie Webbe\inst{1}\fnmsep\thanks{\email{robbie.webbe@irap.omp.eu}}, E. Quintin\inst{2}, N. A. Webb\inst{1}, Gabriele Ponti\inst{3,4,5}, Tong Bao\inst{3}, Chandreyee Maitra\inst{4,6}, Shifra Mandel\inst{7}, Samaresh Mondal\inst{8}}

   \institute{Institute for Research in Astrophysics and Planetology (IRAP), CNRS, Toulouse, 31400, France
    \and
    European Space Agency (ESA), European Space Astronomy Centre (ESAC), Camino Bajo del Castillo s/n, 28692 Villanueva de la Cañada, Madrid, Spain
    \and
    INAF – Osservatorio Astronomico di Brera, Via E. Bianchi 46, 23807 Merate (LC), Italy
    \and
    Max-Planck Institut für extraterrestrische Physik, Giessenbachstrasse, 85748 Garching, Germany
    \and
    Como Lake Center for Astrophysics (CLAP), DiSAT, Università degli Studi dell'Insubria, via Valleggio 11, I-22100 Como, Italy
    \and
    Inter-University Centre for Astronomy and Astrophysics (IUCAA), Ganeshkhind, Pune 411007, India
    \and
    National Science Foundation Fellow, Columbia Astrophysics Laboratory, Columbia University, New York, NY 10027, USA
    \and
    Department of Astronomy, University of Illinois, 1002 W. Green St., Urbana, IL 61801, USA}
   \date{}

  \abstract
   {The study of astronomical transients at high energies provides insights into some of the most extreme physical events in the universe; however, carrying out their detection and fast follow-up studies are limited by instrumental constraints. Search for Transient Object in New observations using Known Sources (STONKS) is a near-real-time transient detection system for \textit{XMM-Newton} offering the capability to detect transients in \emph{XMM-Newton} observations at fainter fluxes than can be achieved with wide survey instruments.}
   {We present the transients detected with the STONKS pipeline found in an \emph{XMM-Newton} multi-year heritage survey of the Galactic plane to identify and classify highly variable X-ray sources that have recently been reported in this region.}
   {We examined the alerts created by the STONKS pipeline from over 200 XMM observations of the Galactic plane, screening for instrumental effects. The 78 alerts associated with 70 real astrophysical sources were then subjected to further temporal and spectral analysis.}
   {From the 70 sources we identified, we were able to classify 32 with a high degree of confidence, including 7 X-ray binaries, 1 $\gamma$-Cas analogue, and 1 magnetar candidate. Of the 70 sources, 23 were detected for the first time in X-rays.}
   {This systematic analysis of publicly available data has shown the value and potential of STONKS in the application to \emph{XMM-Newton} observations. It will enable the community to detect transient and highly variable sources at fainter fluxes than with any other X-ray transient detection systems.}
   \keywords{X-rays: general -- Methods: data analysis}
\titlerunning{STONKS X-ray transients in the Galactic plane}
\authorrunning{Robbie Webbe et al.}
\maketitle
\vspace{-2.5cm}

\section{Introduction}
\label{sec:intro}

The X-ray sky is rich in variable phenomena \citep[e.g.][]{li_populations_2022,khan_exod_2025}, offering many opportunities to probe some of the most energetic astrophysical events. To acquire the energy for X-ray emission, in a variety of sources including stars, stellar remnants, and active galactic nuclei (AGNs), we require extreme physical conditions. These often include strong magnetic or gravitational fields, or very high temperatures. Instabilities in these environments can invoke temporal and spectral changes that offer insights into the processes taking place. Examples include X-ray emission from coronal emission in stars and how it relates to magnetic reconnection events \citep[e.g.][]{gudel_x-ray_2009}, tidal disruption events and how they describe the strong gravity experienced by stars approaching massive black holes \citep[e.g.][]{gezari_tidal_2021}, and Type I X-ray bursts, along with how they relate to thermonuclear burning on neutron stars and the effect on their atmospheres \citep[e.g.][]{lewin_x-ray_1993}.

Several recent X-ray observatories have been designed for the detection of variable X-ray sources and transients, including \emph{Neil Gehrels Swift} Observatory \citep{gehrels_swift_2004}, the Space-based multi-band astronomical Variable Objects Monitor \citep[SVOM; e.g.][]{atteia_svom_2022}, and the \emph{Einstein} Probe \citep{yuan_einstein_2022}. These missions typically employ wide fields of view to maximize the likelihood of detecting transient events. As part of the XMM2Athena project \citep{webb_xmm2athena_2023}, we have developed a near real-time transient detection system for the \textit{XMM-Newton} pipeline. Search for Transient Object in New observations using Known Sources  \citep[STONKS:][]{quintin_stonks_2024} automatically detects any long-term variability within the field of view by comparing the new observation to several available sources of archival X-ray data. STONKS presents an alternative to wide field-of-view missions with high-flux detection thresholds. The long exposure times of pointed \emph{XMM-Newton} observations enable  observers to identify sources that are significantly variable with fluxes down to $3.1\times10^{-16}$\,erg\,cm$^{-2}$\,s$^{-1}$ in the range of 0.5-2.0\,keV for exposures of $\sim$100\,ks \citep{hasinger_xmm-newton_2001}.  These parameters are several orders of magnitude below those available to survey missions. As of the AO24 proposal cycle, STONKS is now being applied to all \textit{XMM-Newton} observations, with alerts shared with principal investigator and (with their prior approval) with the community. This paper presents the first results from testing the pipeline on publicly available data, notably the 2024 \textit{XMM-Newton} Galactic Plane Survey \citep[XGPS: e.g.][]{mondal_intermediate_2022,mondal_discovery_2023,anastasopoulou_study_2023,mondal_periodicity_2024,mondal_xmm-newton_2024}. A survey of the Galactic plane enables us to detect a wide variety of X-ray sources, including objects corresponding to all points of stellar evolution. Here, we present some examples of the highly variable sources we could expect STONKS to detect in the Galactic plane.

Main sequence stars emit X-rays as the result of magnetic reconnection events in their coronae \citep[e.g.][]{favata_stellar_2003,gudel_x-ray_2009} driven by their rotation. Flares from these reconnection events are on timescales of the order of hours, while much longer (i.e. year-long) dynamo-cycles can also affect the X-ray emission by orders of magnitude \citep{orlando_sun_2001}. The rotation periods can be significantly different for stars of the same mass when young \citep{affer_rotation_2013}, but are expected to converge on values determined by their age and mass for all but some of the smallest M-dwarfs \citep{irwin_angular_2011}. The relations between the stellar rotation, bolometric luminosity, and X-ray luminosity are well constrained for different classes of stars \citep{pallavicini_relations_1981}, with a maximum such that log$(L_X/L_{\text{Bol}}) \sim -3$ \citep{vilhu_nature_1984} even for the fastest rotators. Flares, with durations of the order of hours, can significantly affect the luminosity of these objects, emitting total energies of the order of $E_X \geq 10^{32}$\,erg \citep[e.g.][]{pandey_study_2008,pye_survey_2015}. These detections of flares are too short-lived to allow for follow-ups, but they will help identify stellar X-ray flares and help improve our understanding of the phenomenon.

Young stellar objects (YSOs) also display significant X-ray variability as a result of magnetic activity \citep{montmerle_magnetic_1993}. This coronal activity is similar to the activity seen in more evolved main sequence stars, but there are also features in their X-ray emission that may be due to shocks from accretion or outflows, or fluorescence from an irradiated disc \citep{gudel_x-ray_2009}. The X-ray emission from such activity can reach several orders of magnitude higher than that seen from main sequence stars \citep{feigelson_high-energy_1999,stelzer_x-ray_2017}. As this emission is not the result of constant processes, large variations in the X-ray luminosity can be observed, from below detection limits up to luminosities in the range $L_X\sim10^{29-34}$\,erg\,s$^{-1}$ \citep[][etc.]{preibisch_origin_2005,winston_chandra_2018}. These flares have typical rise times on scales of minutes to hours and then decay over several hours \citep{montmerle_magnetic_1993}.

The remnants of stars can also display significant X-ray variability. Cataclysmic variables (CVs) are powered by accretion onto white dwarfs often from Roche-lobe filling K and M-type stars. They can display significant short and long-term variability. Many CVs will display pulsations caused by the rotation of the accreting white dwarf \citep[e.g.][]{kuulkers_x-rays_2006,mukai_x-ray_2017}, but these may be averaged out over the course of typical X-ray observations. More significant (and longer lasting) variability can be observed in CVs over several orders of magnitude due to changes in the accretion flow on to the white dwarf through the disc. For CVs with lower or no magnetic fields (non-magnetic CVs), the disc instability model \citep[e.g.][]{lasota_disc_2001} describes how accretion rate onto the white dwarf can significantly increase. In quiescence CVs have X-ray luminosities of the order $L_X = 10^{29-34}$\,erg\,s$^{-1}$\citep[][and others]{mukai_x-ray_2017,schwope_first_2024}, but can exceed $L_X\geq 10^{34}$ erg s$^{-1}$\citep{kuulkers_x-rays_2006} at their most luminous. These novae \citep[e.g.][]{hameury_dwarf_2017} are characterised by durations of the order of days and can recur with periods of weeks or months, making them ideal for detection by STONKS and follow-up. In magnetic CVs, the magnetic field truncates the inner part of the accretion disc (intermediate polars) or prevents the disc from forming (polars). Thus, variability is due to variations in the mass transfer rate from the donor, $\dot m.$ In cases where white dwarfs exist in binaries with higher mass companions (i.e. symbiotic stars), the accretion rate (and, thus, the luminosity) can vary over several orders of magnitude as the result of variable winds from a high-mass companion \citep{kuulkers_x-rays_2006,mukai_x-ray_2017}.

X-ray binaries (XRBs) are powered by accretion onto stellar mass black holes or neutron stars and they can be similarly variable to CVs. Binaries fed by low-mass donors can display highly variable outbursts \citep[e.g. ][]{remillard_x-ray_2006,bahramian_low-mass_2023} in accordance with the disc instability model. Luminosity changes can encompass several orders of magnitude during these outbursts. Typical quiescent X-ray luminosities lie in the range $L_X \sim10^{30.5-33.5}$, and can surpass $L_X \geq 10^{36}$ at the peak \citep{remillard_x-ray_2006} of an outburst. X-ray binaries with high mass donors can also display large variability as a result of the accretion of variable winds from the high mass companion \citep{fornasini_high-mass_2023}. The exact nature of the evolution of accretion discs during these outbursts and how the hard X-ray producing regions and radio jets are affected are issues that are still poorly understood. Rigorously detecting and monitoring outbursts in XRBs and the appearance and disappearance of jets \citep[e.g.][]{gallo_black_2005,kylafis_formation_2012} associated with these outbursts can further deepen our understanding of these sources. As outbursts can last for several weeks or months, they represent a prime target for detection by STONKS, allowing for prompt monitoring and enabling the detection of significant numbers of new transient XRBs.

Neutron star-powered X-ray sources can, however, also display significant shorter term variability. Type-1 X-ray bursts \citep[][]{lewin_x-ray_1993,int_zand_searching_2019}, caused by runaway thermonuclear burning on the surfaces of neutron stars, have durations of the order of $\sim10-100$\,s. In addition, they can be Eddington-limited with peak luminosities of $L_X\sim10^{38}$\,erg\,s$^{-1}$ \citep[e.g.][]{pastor-marazuela_exod_2020}. Repeated bursts from XRBs powered by accretion on to neutron stars from a high mass companion, namely, a supergiant fast X-ray transient \citep[][]{sidoli_supergiant_2017}, can result in variability by greater than one order of magnitude. 

Neutron stars with very strong magnetic fields ($B \gtrsim 10^{13}$\,G; e.g. \citealt{olausen_mcgill_2014,kaspi_magnetars_2017}), known as magnetars, have quiescent X-ray luminosities in the range $L_X \sim 10^{31-35}$; however, they can display further variability as a result of those magnetic fields. Bursts in magnetars \citep[e.g.][]{gogus_statistical_2000,van_der_horst_sgr_2012} have timescales of under 1\,s, but can reach luminosities of up to $10^{43}$\,erg\,s$^{-1}$ \citep{kaspi_magnetars_2017}. Giant flares are even more energetic with durations on the order $\sim100$\,s and luminosities up to $L_X\sim10^{47}$\,erg\,s$^{-1}$ \citep[e.g.][]{cline_detection_1980,hurley_giant_1999}. While these phenomena are too fast to allow for follow-up after detection by STONKS, the detection of higher flux states could enable the identification of more examples of this class, which would then be monitored for further variability. Outbursts in magnetars, with peak luminosities of the order $L_X\sim10^{36}$\,erg\,s$^{-1}$ \citep{rea_magnetar_2011,coti_zelati_systematic_2018}, tend to be followed by a decay in their luminosity on the order of weeks or months. Therefore, follow-ups are possible if they are propmptly identified through an alert issued by STONKS. 

In this analysis, we present the results of an application of the STONKS pipeline to a recently completed survey of the Galactic plane, illustrating its potential in the near-real time detection of X-ray transients. In Sect. \ref{sec:methods}, we outline the approach that will be taken in processing these survey observations and classifying the STONKS alert sources. In Sect. \ref{sec:results}, we present the results of applying this methodology to the survey observations. In Sects. \ref{sec:discuss} and \ref{sec:sosi}, we discuss how this analysis affects our understanding of variable sources in the Galactic plane and we examine a subset of these sources in greater detail.

\section{Methods}
\label{sec:methods}

\subsection{The \emph{STONKS} pipeline}
\label{subsec:methods-pipeline}

This analysis employs the Search for Transient Objects in New detections using Known Sources (STONKS) pipeline developed by \citet[henceforth, Q24]{quintin_stonks_2024}. This pipeline is based upon a multi-mission catalogue of X-ray sources compiled by cross-matching catalogues of X-ray sources observed by \emph{XMM-Newton} \citep{webb_xmm-newton_2020}, \emph{Swift} \citep{evans_2sxps_2020}, \emph{Chandra} \citep{evans_chandra_2024}, ROSAT \citep{boller_second_2016}, and eROSITA \citep{salvato_erosita_2022,merloni_srgerosita_2024}. After these sources have been collated, flux estimates are then obtained for all detections of each source in the range from 0.1--12\,keV, using an assumed spectrum of an absorbed power law with parameters $\Gamma=1.7$ and $N_H = 3\times10^{20}\text{cm}^{-2}$ (Q24). These flux estimates can then be used as a baseline to estimate the magnitude of new variability in these sources, or to identify significant past variability. The validity of this estimate is examined in detail in Q24. Upper limits are also obtained for all locations where sources have not been detected previously by \emph{XMM-Newton}, but when they were in the field of view, in both pointing and slew observations. The variability of a new detection of a source is computed pessimistically, following 
\begin{equation}
    V = \frac{F_{\text{new,low}}}{\min(F_{\text{past,up}},\text{UL})}
,\end{equation}
where $F_{\text{new,low}}$ is the newly obtained flux measurement less the $1\sigma$ uncertainty on this measurement, compared against $F_{\text{past,up}}$, an archived flux measurement plus the $1\sigma$ uncertainty, or against UL, which is the $3\sigma$ upper limit on the flux at that position. Variations of this formula are applied in the case of sources with past variability or which have decreased in brightness. To create a rigorous estimate of significant variability, only sources with a variability $V \geq 5$ (Q24: false alert rate $\sim0.6\%$) were used to trigger alerts.
Full details for the creation and validation of the STONKS pipeline can be found in \cite{quintin_stonks_2024}.

\subsection{Observations}
\label{subsec:methods-obs}

\begin{figure*}
\centering
\includegraphics[width=0.8\textwidth]{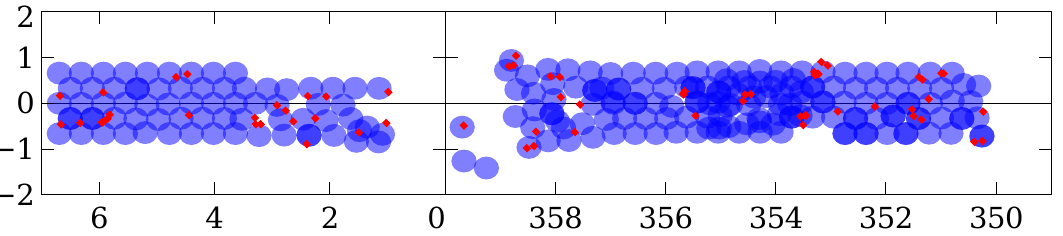}
\caption{Locations of alerts within Multi-Year Heritage Programme observations in galactic coordinates. Observations are presented as blue circles, with radius 15' as an approximation of the footprint of XMM observations and alert sources are indicated by red diamonds.}
\label{fig:alert_locs}
\vspace{-0.5cm}
\end{figure*}

This analysis concerns all of the Multi-Year Heritage Programme observations of the Galactic plane conducted by \emph{XMM-Newton} between 13 March 2021 and 10 October 2024. A total of 231 observations were recorded and processed by the XMM analysis pipeline. Source lists from each observation were then processed by the STONKS pipeline and the alerts were screened in accordance with the methodology outlined in Sect. \ref{subsec:methods-screen}. The pipeline versions used in the processing of observations are listed on the XMM Science Archive\footnote{\url{https://nxsa.esac.esa.int/nxsa-web/\#home}}. The characteristics of these observations are discussed in greater detail in Sect. \ref{subsec:results-genpop} and we show the footprint of the programme in Fig. \ref{fig:alert_locs}.

\subsection{Alert screening}
\label{subsec:methods-screen}

For the majority of observations performed with \emph{XMM-Newton} the source detection pipeline is not affected by observational effects. The sources identified can then be analysed with a high degree of confidence that they do relate to real astrophysical sources. This is reflected by the very high proportion (85\%) of source detections in the most recent version of the XMM Serendipitous Source Catalogue \citep{webb_xmm-newton_2020}, DR14, which are flagged as \texttt{'SUM\_FLAG'} $\leq 1$, indicating a good quality of source detection.

Given the crowded nature of the surveyed region, there is a significant risk that observations may have been affected by very bright sources just outside the field of view of the EPIC camera which can cause the appearance of bright, concentric reflection rings. Once the positions of these spurious "new" sources are compared against archival information they might be wrongly identified as first detections of newly variable bright sources. To mitigate this risk, we manually inspected the observations containing alerts for these and other effects.

High line-of-sight neutral hydrogen column densities in the Galactic plane also create the potential for false alerts. This is due to the spectral assumptions inherent in the STONKS pipeline (Q24), and those of X-ray missions and their catalogues: XMM \citep{webb_xmm-newton_2020}, \emph{Chandra} \citep{evans_chandra_2024}, \emph{Swift} \citep{evans_2sxps_2020}, ROSAT \citep{boller_second_2016,white_wgacat_1994}, and eROSITA \citep{salvato_erosita_2022,merloni_srgerosita_2024}. Flux estimations rely on an assumed spectral shape, relevant for an average point in the sky but typically with a lower column density than in the direction of the Galactic plane, leading to incorrectly derived fluxes. This is further compounded by the overlapping, but not identical, energy ranges when considering observations of the same region by different instruments. Flux estimations from different instruments will be affected to some degree by the spectral assumptions, potentially leading to variable fluxes even for constant sources. We therefore inspected all spectra, after first screening for observational effects, to determine whether any of the alerts were likely to be false alerts due to spectral assumptions.

\subsection{Source classification}
\label{subsec:methods-class}

Once screened, these sources were classified to determine their nature and the underlying phenomena responsible for the variability. This was achieved by examining the features in X-ray data, from new and archival detections and by considering the multi-wavelength information. In the Galactic plane, we expect the majority of alert sources to fall into a small number of categories, associated with Galactic X-ray sources: stars, YSOs, CVs, and XRBs.

The X-ray light curves and spectra were examined using either the pre-created \emph{XMM-Newton} processing pipeline products or by reprocessing the raw event data using \texttt{XMMSAS} version \texttt{xmmsas\textunderscore20230412\textunderscore1735-21.0.0}\footnote{\url{https://www.cosmos.esa.int/web/xmm-newton/sas}}. Source and background regions were extracted for sources on an individual basis to ensure a high signal-to-noise ratio (S/N) and to create source-free background regions from the same CCD, which were at least twice as large as the source regions. Events were extracted for all instruments where the source was detected. Light curves were created with time bins of 1\,s for pn detections and of 5\,s for MOS detections. When more than one spectrum was available for a detection, they were all combined to maximise the S/N. Combined light curves were created using the MOS1 and MOS2 cameras where possible if no pn detection was available. To visually examine time domain variability on a range of scales the light curves were re-binned with time bin sizes of 10\,s, 50\,s, 250\,s, and 1\,ks. Where appropriate, we also attempted to identify periodicities in source event lists using epoch folding and $Z^2_n$ tests contained in the \texttt{stingray} \citep{huppenkothen_stingray_2019} module for \texttt{Python}. X-ray spectra were fit using the \texttt{PyXspec} implementation of \texttt{xspec} \citep{arnaud_xspec_1996}.

Multi-wavelength information was obtained from the SIMBAD\footnote{\url{https://simbad.cds.unistra.fr/simbad/}} Astronomical Database and the VizieR\footnote{\url{https://vizier.cds.unistra.fr/viz-bin/VizieR}} library of astronomical catalogues. The STONKS pipeline queried the SIMBAD database for known sources within 10" of the X-ray source position. The X-ray source locations were also, independent of any SIMBAD matches, queried through VizieR to obtain optical counterparts in the \emph{Gaia} DR3 catalogue \citep{gaia_collaboration_gaia_2023}, near-infrared (NIR) counterparts in the 2MASS All-Sky Catalog of Point Sources \citep{cutri_vizier_2003}, and mid-IR (MIR) counterparts in the AllWISE data release \citep{cutri_vizier_2021}.

Finally, we also considered previous classifications of X-ray sources in the \emph{XMM-Newton} and \emph{Chandra} archives using machine learning. We searched for previously detected sources in the 4XMM-DR12 catalogue\footnote{\url{http://xmmssc.irap.omp.eu/Catalogue/4XMM-DR12/4XMM_DR12.html}} \citep{webb_xmm-newton_2020} that were classified using a naive Bayes classifier by CLAXBOI \citep{tranin_probabilistic_2022} and those sources with previous \emph{Chandra} detections in the \emph{Chandra} Source Catalog Release 2.0\footnote{\url{https://cxc.harvard.edu/csc2.0/}} \citep{evans_chandra_2024} as classified by MUWCLASS \citep{yang_classifying_2022}. We then collated this information and applied the following classification methodology:
\begin{itemize}
\item Verify the location of any STONKS-identified SIMBAD counterpart is coincident with the X-ray source.
\item Inspect light curves for distinctive patterns of variability.
\item Perform fits to the X-ray spectra using absorbed (\texttt{tbabs * }) blackbody (\texttt{bbody}), powerlaw (\texttt{powerlaw}), plasma (\texttt{apec}), and thermal bremsstrahlung (\texttt{bremss}) models.
\item Identify optical and IR counterparts to the X-ray sources, consider the colour differences, and determine the X-ray-to-optical flux ratios or the limits on the ratio when optical counterparts are not present.
\item Inspect the light curves and X-ray spectral fits to any previous XMM observations of the source.
\item Identify source classifications from the CLAXBOI and MUWCLASS catalogues.
\end{itemize}

These steps are not considered hierarchically and all available information forms part of the overall classification. Additional steps which were taken to confirm membership for each of the individual source classes are detailed in Appendix \ref{app:class-criteria}. For those sources that remained unclassified at this stage we then made provisional classifications as to their nature based on the available information. These should, however, be treated with the utmost caution and require future follow-up.

\section{Results}
\label{sec:results}
\subsection{General population}
\label{subsec:results-genpop}

Between the 13th March 2021 and 10th October 2024 a total of 231 observations were conducted as part of an \emph{XMM-Newton} Multi-Year Heritage Project survey of the Galactic plane. For 18 observations, no source detection was performed by the \emph{XMM-Newton} pipeline as the exposure time following screening for background flaring was below the limit of 10\,ks. The remaining 213 were processed by STONKS and 71 triggered alerts. The durations of these 213 observations ranged from 11.5\,ks to 48.4\,ks.

Over the duration of the survey, this corresponded to an average of 0.11 alerts per day just from these survey observations, with an expectation $E(\text{Alert}) = 0.62$ alerts per observation. With a total observation time across the survey of 4380\,ks, the average alert rate was 0.032 alert\,ks$^{-1}$. 

There were a total of 142 alerts from the 231 observations. These were screened according to the procedure described in Sect. \ref{subsec:methods-screen}, while 64 were rejected as observational artefacts. All of these spurious alerts were for first detections of new sources, and 63 also triggered alerts that the spectrum could not be well described using standard assumptions. The remaining 78 alerts, summarised in Table \ref{tab:results-alertsummary}, were examined further, with 54 of them also triggering spectral alerts. These 78 alerts were associated with 70 sources, with 4 sources triggering 2 alerts, and 2 sources triggering 3 alerts. Of these 78 alerts, 6 were expected to be caused spuriously by issues with flux estimations caused by the extreme spectra observed and a further 3 were deemed questionable. We included these nine alerts, which may be the result of spectral assumptions, nonetheless, in our subsequent analysis as their variability cannot be conclusively excluded without a more detailed spectral analysis of all multi-observatory data.

\begin{table}
\caption{Summary of the 142 STONKS alerts triggered during the Galactic Plane survey.}
\label{tab:results-alertsummary}
\centering
\begin{tabular}{lcc}
\hline\hline
Alert class & True alerts & Instrumental alerts \\
\hline
All & 78 & 64 \\
\hline
First detection & 25 & 64 \\
High-flux state & 29 & 0 \\
Low-flux state & 15 & 0 \\
Past variability & 9 & 0 \\
\hline
\end{tabular}
\tablefoot{True alerts are those defined as not being contaminated by artefacts as described in Sect. \ref{subsec:methods-screen}.}
\end{table}

Alert Source 20 (SIMBAD association:\ cataclysmic binary V478 Sco) triggered two alerts: one for past variability and one for a low flux state. It was the only such to trigger two alert classes. All other sources with multiple alerts triggered alerts of the same class. One source that triggered multiple alerts did not also trigger alerts for extreme spectra in all of those alerts. Alert Source 51 (SIMBAD association - V* V1017 Sco) triggered three alerts for a low flux state, but one of those three did not have an alert for an extreme observed spectrum. When considering the locations of sources within the Galactic plane (see Fig. \ref{fig:alert_locs}), there is no apparent large-scale structure or overdensity, although some local clustering of alerts might be present.

The fluxes of the alert sources cover more than three orders of magnitude. The faintest source was associated with a low flux state alert at $5.1\times10^{-15}$ erg\,s$^{-1}$\,cm$^{-2}$, while the brightest source was associated with a high flux state alert at $9.7\times10^{-12}$ erg\,s$^{-1}$\,cm$^{-2}$. The median and mean alert fluxes were $2.44\times10^{-13}$ erg\,s$^{-1}$\,cm$^{-2}$ and $2.23\times10^{-13}$ erg\,s$^{-1}$\,cm$^{-2}$, with the 5-95\% interval for fluxes ranging from $0.26 - 24.66 \times10^{-13}$ erg\,s$^{-1}$\,cm$^{-2}$. With the exceptions of the brightest sources these fluxes are below the detection limits of other X-ray observatories focussed on the identification of variable sources, for example MAXI.

\subsection{Multi-wavelength counterparts to X-ray alert sources}
\label{subsec:results-mwcparts}

The 70 Alert Sources were cross-matched with the \emph{Gaia}, 2MASS, and AllWISE source catalogues as described in Sect. \ref{subsec:methods-class}. Of these, 25 (35.7\%) had SIMBAD associations as per the STONKS pipeline. The classifications of the 25 sources with SIMBAD associations are 11 stars (including high proper motion, Be, RS Cvn variable), 5 YSOs, 2 X-ray sources, 2 XRBs, 2 IR sources, 2 radio sources, and 1 CV. A significant minority (44\%) of these SIMBAD associations are for stars, with a small number of others in expected Galactic transient classes. Those sources classed as X-ray, IR or radio sources (6 of 25) do not have further classifications with specific phenomenologies, and only one of the IR sources (source 13) also had a prior X-ray detection noted.

\begin{figure}
\centering
\includegraphics[width=\columnwidth]{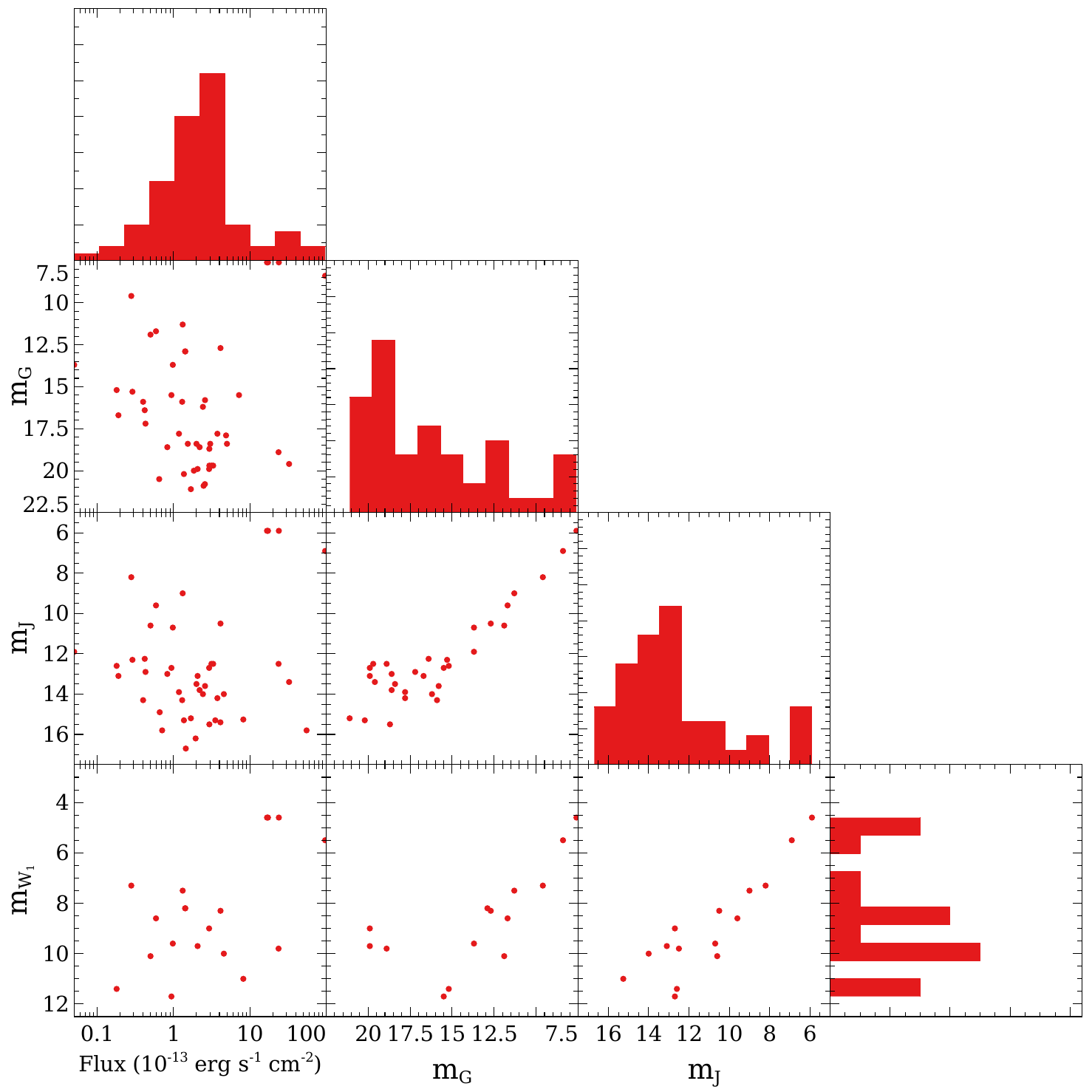}
\caption{X-ray fluxes of alert sources and apparent optical, NIR, and MIR magnitudes for the associated counterparts of these sources.}
\label{fig:res-alerts-mwcounts}
\vspace{-0.25cm}
\end{figure}

We used  \emph{Gaia}, 2MASS, or AllWISE to identify  counterparts for  49 (70\%) of the 70 alert sources, with 35 of those having a counterpart in at least two bands. The numbers of sources with counterparts are detailed in Table \ref{tab:results-mwsummary}. There are fewer  sources (13) that display a cross-match in all three bands than sources that do not display any match; there is a significant minority of 21 (30\%) that have no identifiable optical or IR counterpart. However, this was not an unexpected outcome considering the target area of the survey, which is prone to high levels of optical extinction. More than half of our alert sources were associated with an identified optical source or near IR source, and just under half of the sources have both a \emph{Gaia} and a 2MASS counterpart, allowing for a colour analysis. The mean X-ray source localisation was \ang{;;0.9} and within 3$\sigma$ ranges, there were only eight sources with more than one potential optical counterpart. In the few cases where multiple counterparts appeared equally likely, considering distances weighted by positional uncertainties, the results of the spectral fitting and temporal examination formed part of the decision as to the likely counterpart. For the purposes of this study, we did not consider the likelihood of spurious multi-wavelength matches; however, we did not classify sources solely on the basis of their multi-wavelength properties; hence, a small number of coincidently overlapping sources should not significantly affect our aggregate results. In Fig. \ref{fig:res-alerts-mwcounts} we show the distributions of X-ray fluxes for alert sources against their associated optical, NIR or MIR counterpart magnitudes. Full listings of STONKS X-ray alert sources and their multi-wavelength counterparts are detailed in Table \ref{tab:app_screened_alerts_mw1}.

\begin{table*}
\centering
\caption{Catalogued and multi-wavelength counterparts to STONKS X-ray alert sources per their alert classes.}
\label{tab:results-mwsummary}
\begin{tabular}{lcc|ccc|ccc|cc}
\hline\hline
Alert class & No. & SIMBAD & Opt. & NIR & MIR & Opt \& NIR & Opt \& MIR & NIR \& MIR & Opt, NIR \& MIR & None \\
\hline
All & 70 & 25 & 40 & 40 & 16 & 31 & 14 & 15 & 13 & 21 \\
\hline
First det. & 23 & 4 & 11 & 8 & 3 & 5 & 1 & 3 & 1 & 9 \\
High flux & 27 & 10 & 11 & 13 & 4 & 9 & 3 & 4 & 3 & 12 \\
Low flux & 12 & 8 & 12 & 11 & 8 & 11 & 8 & 7 & 7 & 0 \\
Past var. & 9 & 4 & 7 & 9 & 3 & 7 & 3 & 3 & 3 & 0 \\
\hline
\end{tabular}
\tablefoot{The SIMBAD column indicates those sources which have SIMBAD counterparts as per the STONKS pipeline, but does not consider the quality of the association. The Opt., NIR, and MIR columns lists those sources with \emph{Gaia}, 2MASS, or AllWISE counterparts. The last five columns list the numbers of sources with combinations of these counterparts.}
\end{table*}

\begin{figure*}
\centering
\includegraphics[width=\textwidth]{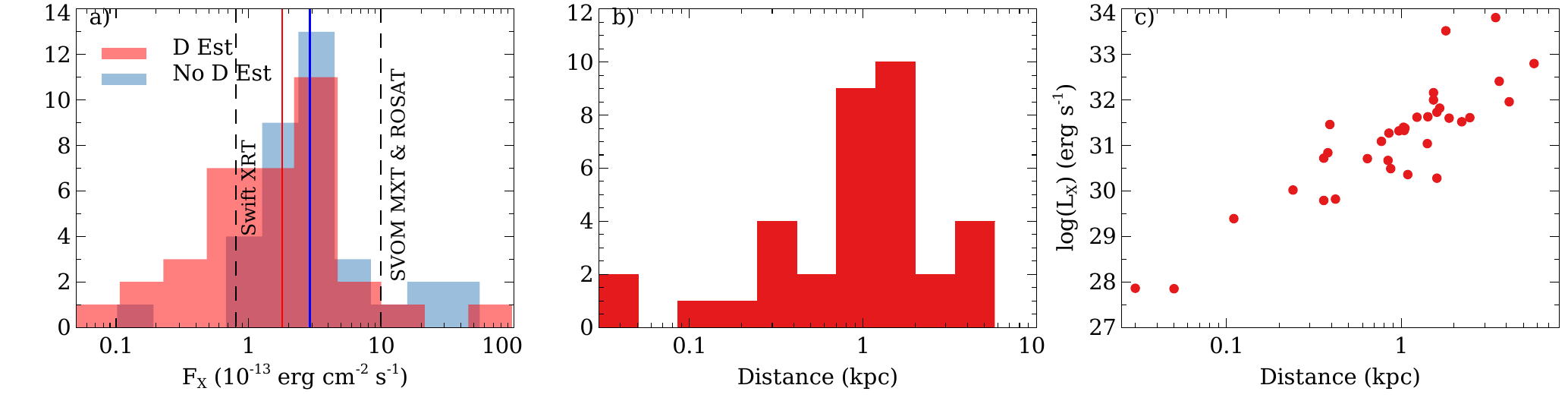}
\vspace{-0.5cm}
\caption{Distributions of X-ray fluxes, distance estimates, and derived luminosities for STONKS alert sources. Panel (a): Comparison of flux distributions for sources with distance estimates (red), and without (blue).\ Panel (b): Distance estimates for the 35 sources where one is available. Panel (c): Distribution of distances and X-ray luminosities for these 35 sources. In panel (a) we also display the 10\,ks sensitivity limits for source detection in some other X-ray transient detectors.}
\label{fig:res-dist-lx}
\vspace{-0.5cm}
\end{figure*}

Distance measurements were available for 35 of the alert sources. These distance estimates covered more than two orders of magnitude, from 28\,pc to 5.76\,kpc. For those sources without an estimated distance (as there is no optical counterpart), we considered the possibility that the lack of a visible counterpart is due to high absorption along the line of sight. In these cases we use an estimate of 8\,kpc to derive X-ray luminosities, as the approximate distance to the Galactic centre. This will provide an appropriate upper estimate for the luminosities of potentially highly absorbed sources which would explain the lack of an optical or IR counterpart. Luminosity estimates for these sources can then easily be scaled to consider their brightness at greater or lesser distances from this benchmark on an individual basis, as advised by the results of temporal and spectral analysis. In Fig. \ref{fig:res-dist-lx}, we show the distributions of X-ray fluxes for sources with and without distance estimates, along with the distances and luminosities derived for sources with estimated distances.

There is a clear overlap between the two populations, but we find that the average flux for those sources without distance estimate is non-significantly higher (see Fig. \ref{fig:res-dist-lx}). For sources with an optical counterpart and distance estimate, the mean flux leads to $log(F_X)= -12.75\pm0.61$, while for those without, we have $log(F_X)= -12.54\pm0.51$.

\subsubsection{Firm classification of sources}
\label{subsec:results-firmclass}

The 70 X-ray selected sources were classified using the methodology outlined in Sect. \ref{subsec:methods-class} and Appendix \ref{app:class-criteria}. These classifications allowed us to identify 38 sources with confidence:
21 stars (including 1 RS CVn star), 7 XRBs (including 1 HMXB), 4 CVs, 4 YSOs, 1 $\gamma$-Cas analogue, and 1 magnetar candidate.
Of these, 13 stars (5 XRBs, 3 CVs, 1 YSO, and the magnetar candidate) had not been previously identified as such. They either had no classification, an alternate classification, or had been identified using a single wavelength-band detection only. These firm classifications   based on our methodology are listed in Table \ref{tab:app_all_class1}.

For 3 of the 25 STONKS-selected sources (sources 1, 2, and 18) the SIMBAD associations automatically generated by the pipeline were not appropriate to the X-ray source location. Some source confusion in crowded regions, such as the galactic plane is not unexpected due to the relatively large error circle used to search for SIMBAD counterparts  as per the STONKS pipeline (\ang{;;10}: Q24). In the cases of sources 1 and 2, our classification identified different types to those in SIMBAD. For four further sources (4, 13, 23, and 57) we were able to provide firm phenomenological classifications. These sources had previously only been identified in SIMBAD using multi-wavelength photometry.

\subsubsection{Provisional classification of sources}
\label{subsec:results-provclass}

Following the classification of the alert sources, there were 32 sources remaining that could not be classified with an appropriate level of confidence. We re-applied the same methodology as previously, but relaxed the limits for classes on the optical-to-X-ray flux ratio (or colour comparisons) to find the best combination of multi-wavelength, temporal, and spectral conditions to provide a provisional classification. These should be treated with great caution, but may be used as a starting point for future analysis. This provisional classification allows for the identification of a further 12 XRBs, 8 stars, 7 CVs, and 5 YSOs. 

\section{Discussion}
\label{sec:discuss}

\subsection{Alert source characteristics}
\label{subsec:discuss-alert-characteristics}

We found associations (p<0.2\%) between the type of STONKS alert and the existence of multi-wavelength counterparts for the sources in our survey. We found $\chi^2$ values of 14.4, 18.2, and 14.6 for three degrees of freedom for associations of alert type with optical, NIR, and MIR counterparts, respectively. The association between alert type and the existence of an NIR counterpart is more significant (3$\sigma$) than in the optical and MIR bands. In all cases we find that the 'low flux' and 'past variability' alerts are more likely to be associated with multi-wavelength counterparts. These associations are unsurprising given that 10 of the 12 sources that triggered low flux alerts are classified as stars, while the remainder can be classified as CVs. In both cases, the expected reddening-corrected optical-to-X-ray flux ratios \citep{pye_survey_2015,schmitt_nature_2022} indicate that these sources would be expected to have multi-wavelength counterparts. The majority of sources that triggered past variability alerts were also classified as stars or YSOs. Again, we would expect to see this association as a result of the expected optical and IR-to-X-ray flux ratios and how bright they would expect to appear in optical and IR bands.

We note that  sources identified as CVs in our sample tend to be located at greater distances than those from two optically selected samples \citep{ramsay_distances_2017,inight_catalogue_2023}. For the six sources in our sample that have distance estimates, we found an average distance (and standard deviation on such) of $d=1.43\pm0.96$\,kpc, with the other five being too faint optically to have a reliable parallax measurement. A \emph{Gaia}-selected sample of 16 CVs had a significantly lower average distance $d=0.30\pm0.17$\,kpc \citep{ramsay_distances_2017}, while a much larger sample of 507 CVs identified by the \emph{Sloan Digital Sky Survey} had a non-significantly lower distance distribution, $d=1.03\pm0.95$\,kpc \citep{inight_catalogue_2023}. With more than half of the CVs identified by our analysis having no reliable distance measurement, potentially being too distant for counterparts to be detected, STONKS could be identifying a significantly more distant sample of CVs than those found in optical surveys. More complete populations of CVs out to greater distances are key to constraining their number densities and so stellar evolution pathways.

\subsection{X-ray source populations}
\label{subsec:discuss-x-ray-pops}

Previous XMM surveys of the Galactic plane have considered overall source populations in this region, and searches for periodically variable sources \citep{mondal_periodicity_2024}, but none have specifically targeted highly variable sources. Additionally, the surveys of \cite{motch_source_2003,motch_x-ray_2010} and \cite{nebot_gomez-moran_xmm-newton_2013} examined low-galactic-latitude regions. In all previous cases a significant number of sources remained unidentified, and so we compare the identified sources with the 38 for which we provided firm classifications. In all cases, the greatest proportion was for stellar sources, with at least 70\% of the identified sources in each case and up to 92\% in the case of \cite{motch_source_2003}. Stars were also the greatest class in our sample, albeit for a much smaller proportion (55\%). As the most populous sources in the Galaxy, they ought to comprise the largest class and the marginally lower proportion in our sample could be representative of the smaller magnitude of their variability compared to other X-ray sources. The observational characteristics may also bias STONKS towards triggering alerts for harder X-ray sources in this region. With no alerts created for non-detections of previously identified sources, the relatively low average flux of sources identified as stars ($log_{10}(F_X)=-12.9$) is likely a contributing factor. If some sources were in low flux states during the survey then they could be too faint to be detected and compared with previous detections. By contrast, the number of alerts for compact object powered sources (WD, NS, or BH) is greater (34\%) than in any previous survey -- 8\% \citep{motch_source_2003}, 30\% \citep{motch_x-ray_2010}, and $\sim$1\% \citep{nebot_gomez-moran_xmm-newton_2013}; in addition, the average flux is significantly greater for these sources ($log_{10}(F_X)=-12.2$). This ability to detect compact object-powered sources through variability was also identified by \citet{lin_classification_2012}. That the difference is greatest between our sample and that of \cite{nebot_gomez-moran_xmm-newton_2013} is striking given that it is closest in the number of fields considered, and observation duration. However, the fact that the \cite{nebot_gomez-moran_xmm-newton_2013} survey considers a much greater range of galactic longitude could be a significant factor, since we would expect compact object-powered systems to be associated with the most evolved populations of sources. With the recent survey being focussed only on the centre of the Galaxy, we would expect to detect a greater proportion of stellar remnants.

In comparison with other XMM populations of X-ray variability-selected sources, we have again found that they are dominated by stellar sources. The first large scale implementation of the EXOD detection algorithm \citep{pastor-marazuela_exod_2020} considered all observations comprising the 3XMM-DR8 catalogue where the EPIC pn camera was active in full-frame mode. They identified 734 point sources that are not associated with AGNs. Of these, 515 ($\sim$70\%) were associated with stellar sources and 153 ($\sim$21\%) with compact object-powered sources which were not AGNs. The EXOD methodology \citep{pastor-marazuela_exod_2020, khan_exod_2025} detects sources with significant in-observation variability and so, the indications are that both long and short-term variability might be able to detect a greater proportion of compact objects over stellar sources than surveys alone. Unfortunately, the results of \citet{pastor-marazuela_exod_2020} do not delineate between the different classes of compact object and so, we cannot conclude which approach is more appropriate for detecting CVs, neutron stars, or black hole sources. The highly variable sources as identified in the \emph{XMM-Newton} Slew survey (XMMSL2\footnote{\url{https://www.cosmos.esa.int/web/xmm-newton/xmmsl2-ug}}) by \citet{li_populations_2022}, whose methodology is also based on long-term variability, again display a similar split in the numbers of point sources. Of the sources not associated with galaxies, they detected 100 stars and 27 compact object-powered sources from a total of 265 variable sources. As the slew survey covers a large fraction of the entire sky, they found a large number of highly variable sources associated with galaxies, which we would not expect to detect in the survey of the inner Galactic disc analysed by STONKS. Within the population of compact object sources identified by \citet{li_populations_2022} there were 6 CVs (this work: 4) and 21 XRBs (this work: 8), indicating that the Slew may be preferential for the detection of XRBs. This is not unexpected given that the Slew survey has a far higher detection threshold than pointed observations and CV sources have lower luminosities than XRBs in their high states. As such, we would expect the Slew survey to preferentially detect more luminous sources.

A \emph{Swift} survey of the Galactic plane \citep{oconnor_swift_2023} identified 928 sources, of which 447 were identified as variable by their criteria; however, we note that the study was focussed on a wider area of the Galactic plane. From the 928 sources, 73 could be identified and in their sample of identified sources, 48 of them (66\%) were associated with compact object sources, including XRBs, magnetars, and CVs. This is a higher proportion than what is contained in our sample of STONKS-selected sources, but this may be as a result of the observing and classification strategy. The observations comprising this \emph{Swift} survey were no longer than 5\,ks and, as a result, the observed source fluxes end up focussed on brighter sources. In the sample of sources considered by \citet{oconnor_swift_2023} the mean flux in the energy range 0.3--10.0\,keV is $2.9\times10^{-13}$\,erg\,s$^{-1}$\,cm$^{-2}$ and the 5--95\% interval ranges from $0.4-46.8\times10^{-13}$\,erg\,s$^{-1}$\,cm$^{-2}$, which is consistently higher than the same values in this STONKS-selected population for a broader energy range. Additionally, their classification methodology focussed only on sources which have previously been identified in SIMBAD or other catalogues, and is not restricted to the identified variable sources. This will necessarily be biased towards brighter sources with rigorous classifications and with 825 (92\%) of their population, it is not possible to rigorously conclude whether there are significant differences among the two full populations.

\begin{figure}
\centering
\includegraphics[width=\columnwidth]{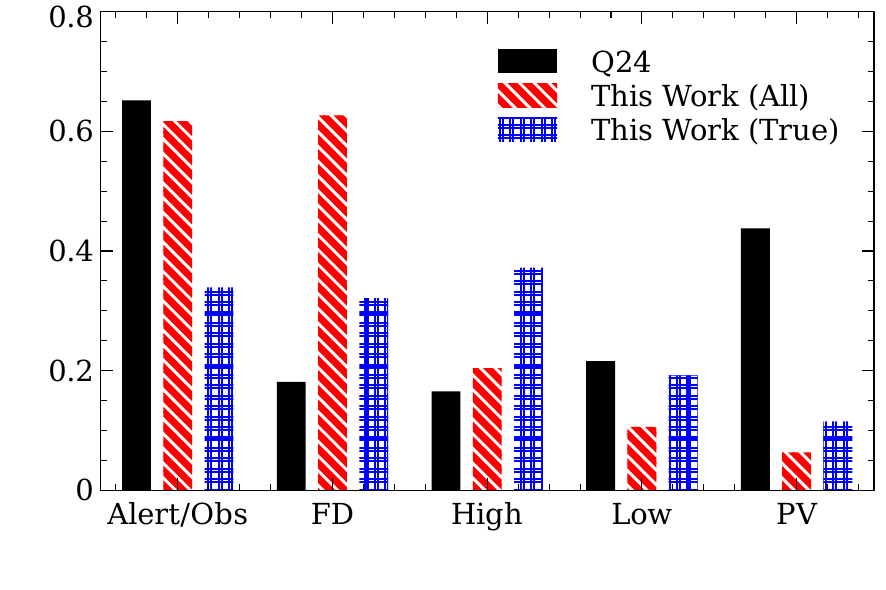}
\vspace{-0.75cm}
\caption{Comparison of alerts and alert types in this work and as per Q24. The first column represents the alerts per observation, and the other four represent the proportion of alerts in each category. The black solid bar represents the results of Q24, while the red diagonally striped bar represents the total alerts and the blue crossed bar show the true alerts obtained during this work.}
\label{fig:alert_pop_change}
\vspace{-0.25cm}
\end{figure}

The only other large-scale implementation of the STONKS pipeline was carried out by \cite{quintin_stonks_2024}, retroactively applying  the transient detection system to all observations of 2021. The total number of alerts being triggered per observation from this MYHP was slightly lower, with 0.617\,alerts per observation (Q24: 0.652). This drops significantly to 0.339\,alerts per observation when we consider that 45\% of the total alerts (Q24: $\lesssim20\%$) were spurious and due to observational effects. We also find a significant difference in the types of alerts triggered, with far more triggers for the first detections and high-flux states: 32.1\% (Q24: 18.1\%) and 37.2\%, respectively (Q24: 16.5\%); as opposed to past variability, 11.5\% (Q24: 43.8\%). We show the differences in these distributions in Fig \ref{fig:alert_pop_change}.

For those MYHP observations where the XMM pipeline performed source detection, the average duration was $\bar{\text{dur}}=19.8$\,ks, with a standard deviation of $\sigma_{\text{dur}}=7.4$\,ks, as opposed to the observations analysed in Q24, where $\bar{\text{dur}}=39.8$\,ks, and $\sigma_{\text{dur}}=29.6$\,ks. The longer average duration during the observations comprising the analysis in Q24 means sources can be detected to a lower flux level. The reduced sensitivity in the MYHP observations reduces the number of sources which can be detected within any given field. Additionally, the high levels of absorption in the direction of the Galactic plane significantly affect the observed flux at lower X-ray energies. This reduces the observed flux from sources which are present and can further reduce the number of sources which are detected and so processed by the STONKS pipeline. The STONKS pipeline currently does not trigger non-detection alerts for sources which have been detected previously. 

The total number of sources detected during the MYHP observations was 18\,445 (Q24: 12\,584). However, a large proportion of these are due to observational artefacts, caused by the high density of bright sources within the Galactic plane. For those observations in this dataset that have been processed and included in the 4XMM DR14 Serendipitous Source Catalogue\footnote{\url{http://xmmssc.irap.omp.eu/Catalogue/4XMM-DR14/4XMM_DR14.html}} (182 of the 213 with source detections), only 7\,397 of the 15\,250 sources have \texttt{'SUM\_FLAG'==0} \citep{webb_xmm-newton_2020}, indicating the highest quality of clean detection. Given that at least half of the detections in this sub-sample of the observations analysed have questionable characteristics, we cannot determine with utmost certainty how many of the total 18\,445 detections are related to true X-ray sources. Without a reliable comparison of true detected sources in the MYHP sample, we can draw no rigorous conclusions as to the intrinsic variability of Galactic plane sources as opposed to a universal population. Across the full population of XMM detections in the DR14 catalogue, more than 73\% (759741 of 1035832) detections have \texttt{'SUM\_FLAG'==0}. We therefore expect that as STONKS is now being applied to the all observations that a greater proportion of new alerts will be due to real variability rather than observational artefacts.

\section{Detailed variable source classifications}
\label{sec:sosi}
Here we further investigate a sample of sources for which there is complementary multi-wavelength data available. We present details of six of these to display the value of STONKS in detecting such objects. These are the alert sources numbered 1, 9, 38, 45, 57, and 61. We address them according to their classifications in the following Sects. \ref{subsec:res-magnetar}, \ref{subsec:res-gamma-cas}, \ref{subsec:res-newxrb}, and \ref{subsec:res-newcv}.

\subsection{Magnetar candidate:\ 4XMM J175136.8-275858}
\label{subsec:res-magnetar}
\begin{figure}
\centering
\includegraphics[width=\columnwidth]{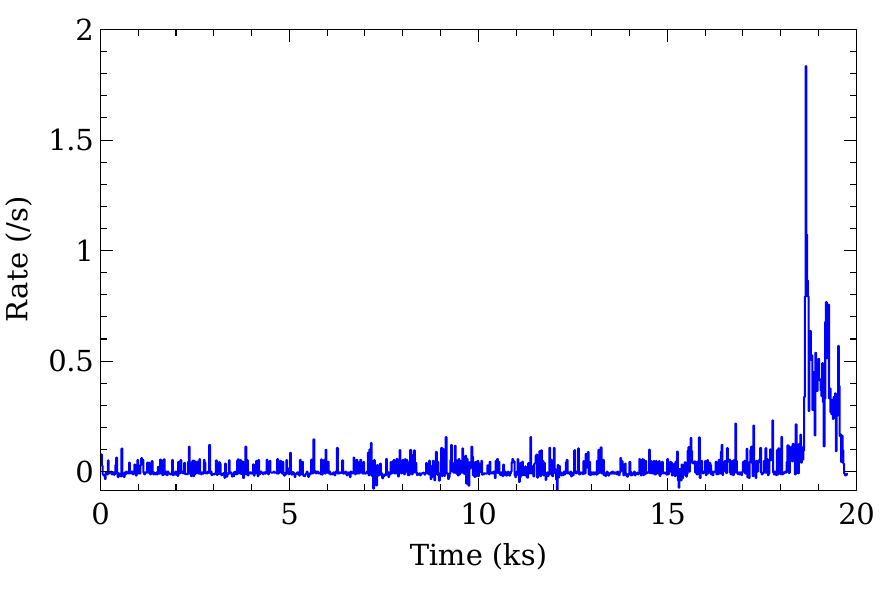}
\vspace{-0.75cm}
\caption{Light curve for source 45, 4XJ1751-2759, during observation 0886121001. The solid line shows the background-subtracted countrate for photons in the energy range 0.2--12.0\,keV. Figure adapted from \citet{webbe_4xmm_2025}.}
\label{fig:4XJ1751-obslc}
\vspace{-0.25cm}
\end{figure}

Source number 45, 4XMM J175136.9-275858 (henceforth, 4XJ1751-2759) was identified as being in a high-flux state in observation 0886121001. An examination of the light curve identified a fast, significant, increase in brightness within the last 5\,ks of the observation, with the increase in count rate coming at photon energies above 2\,keV. We show this in Fig. \ref{fig:4XJ1751-obslc}. This flaring behaviour was also identified during post-processing of X-ray transient candidates identified by the EPIC X-ray Outburst Detector (EXOD; \citealt{pastor-marazuela_exod_2020,khan_exod_2025}), which was developed alongside the STONKS pipeline for the XMM2Athena project \citep{webb_xmm2athena_2023}.

There are two further Galactic plane MYHP observations of the source with \emph{XMM-Newton} within the six months of the observation that triggered the alert, along with three \emph{Chandra} observations with 4XJ1751-2759 in the field of view of during the preceding five months. During one of the \emph{Chandra} observations, 4XJ1751-2759 was not detected with a constraining upper limit. 

During the outburst we find that the flux of 4XJ1751-2759 increases by nearly two orders of magnitude from $3.05\times10^{-13}$erg\,s$^{-1}$\,cm$^{-2}$ to $2.63\times10^{-11}$erg\,s$^{-1}$\,cm$^{-2}$ within a period of a few hours. The flux of 4XJ1751-2759 is still higher than this initial level nearly six months later by a factor of $\sim$2, indicating a long duration for the flux decay. Possibly a greater drop is still to come if the true quiescent flux level is below detectable limits, as implied by the constraining non-detection with \emph{Chandra}.

A spectral fitting to the detection of 4XJ1751-2759 during observation 0886121001 showed that it was well described by a combination of blackbody components, but with $kT > 10$\,keV. We found a blackbody temperature, albeit unconstrained, of 200\,keV during the observed outburst. Fitting to the spectra of the other observations revealed blackbody temperatures consistently between $\sim$1--4\,keV. This value is also consistent with the temperature seen in the known magnetar XTE J1810-197 at a similar time after an observed outburst \citep{borghese_x-ray_2021}. 

During the outburst, the X-ray-to-optical-flux ratio would lead to $log(F_x/F_{Opt}) \sim2.5$ if the detected optical companion were to be associated with the X-ray emission. At the distance of this optical/NIR source, estimated at greater than 5.5\,kpc \citep{bailer-jones_estimating_2021}, this would imply an X-ray luminosity of $L_X \geq 1 \times 10^{35}$\,erg\,s$^{-1}$. This combination of X-ray to optical flux ratio and X-ray luminosity means that the most likely classification for 4XJ1751-2759 is a magnetar. A search for pulsations in the X-ray observations did not return a significant result, but this does not preclude a magnetar classification as they simply may not be detectable in the current observational data. The full details of the temporal and spectral analysis of this source can be found in \citet{webbe_4xmm_2025}.

\subsection{$\gamma$ Cas analogue:\ 4XMM J175328.4-244627}
\label{subsec:res-gamma-cas}

Source 38, 4XMM J175328.4-244627, was identified as being in a high flux state in observation 0886090801 and it was associated with the SIMBAD variable star HD 162718. HD 162718 was identified as a $\gamma$-Cas analogue star by \citet{naze_hot_2018} as a result of its association with a BE star (spectral B3/5), relatively high X-ray luminosity of $-6 < log(L_X/L_{Bol}) < -4$, and spectral features, which included the presence of a thermal component with temperature $kT \geq 5-6$\,keV. It was subsequently identified as periodically variable during a blind search for periodicities in X-ray sources in the Galactic plane by \citet{mondal_periodicity_2024}, with a period of 4.1\,ks. The nature of the X-ray emission from $\gamma$-Cas and its analogues is unknown, with theories citing emission from accretion onto a compact companion and interactions between the decretion disc and the star itself \citep[see ][for a review of these mechanisms]{langer__2020}. HD 162718 was identified as a binary candidate by \citet{naze_velocity_2022} alongside 12 of the 24 other confirmed $\gamma$-Cas sources. The combination of an optically  very bright, giant, young Be companion star with long orbital period would appear to rule out the provisional classification as a polar, by \citet{mondal_periodicity_2024}; however, it could allow for a white dwarf in a symbiotic binary system. The STONKS long-term light curve indicates that this latest observation continues a trend of increasing flux identified in the previous \emph{XMM-Newton} pointed observation in March 2019. The spectral changes indicate that this is, at least in part, a result of increased emission at higher energies. The parallax of the source as per the \emph{Gaia} DR3 catalogue places HD 162718 at a distance of 1.80\,kpc implying a luminosity of $log(L_X) = 33.5$\,erg\,s$^{-1}$. 

\begin{figure*}
\centering
\includegraphics[width=0.9\textwidth]{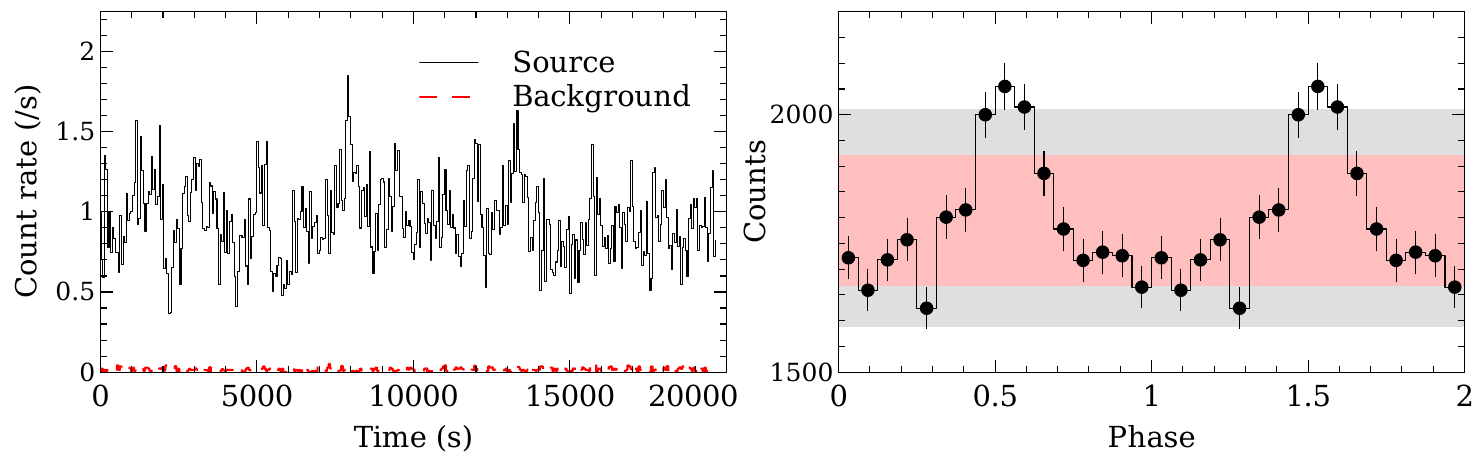}
\vspace{-0.25cm}
\caption{Short-term variability of source 38, 4XJ1753-2446, during observation 0886090801. The left panel displays the background-subtracted 0.2--12.0\,keV light curve with time bins of 50\,s. The source light curve is shown as a solid black line and the background rate is a red dashed line. The right panel shows the light curve folded at a period of 2.03\,ks with time bins of 127\,s. The highlighted red and grey regions show the $3\sigma$ and $5\sigma$ confidence regions, respectively.}
\label{fig:src38-lc-pulse}
\vspace{-0.4cm}
\end{figure*}

An inspection of the light curve identified significant short-term variability. Epoch-folding and $Z^2_n$ searches in the frequency range from $5.0\times10^{-5}-5$\,Hz revealed several significant features. The most significant feature is located at $f=4.9\times10^{-4}$\,Hz for a period of $P \sim2.1$\,ks and is significant at the $5\sigma$ confidence level. This timescale is half of the value reported by \citet{mondal_periodicity_2024} and this might indicate that the pulsation found in that case was double the true period. The light curve and pulse profile are shown in Fig. \ref{fig:src38-lc-pulse}. The pulsations detected in HD 162718 occur on a similar timescale to those observed, albeit transiently, in other $\gamma$-Cas systems, such as $\gamma$-Cas (6.1\,ks and 8.1\,ks; \citealt{frontera_time_1987,haberl__1995}), HD 110342 ($\sim$14\,ks; \citealt{torrejon_bepposax_2001}), and HD 161103 (3.2\,ks; \citealt{lopes_de_oliveira_new_2006}). An examination of the previous pointed \emph{XMM-Newton} observation of HD 162718, observation 0840910501, also identified pulsations at $f=4.5\times10^{-4}$\,Hz for a period $P \sim2.2$\,ks significant at the $5\sigma$ confidence level. 

The combined spectrum from all three instruments was binned for a minimum S/N of 20.0, and fit to absorbed models consisting of several plasma temperature components, and a combination of blackbody and power law components, with or without emission lines. The spectrum was best fit by a combination of an absorbed blackbody and power law component, with temperature of $kT=0.14^{2.95}_{0.03}$\,keV, photon index $\alpha=1.38^{+0.06}_{-0.06}$, an emission line at 6.7\,keV, and with a column density of $N_H = 0.53^{+0.14}_{-0.10}$\,cm$^{-2}$. All the spectral fits to the observation of 4XMM J175328.4-244627 are given in Table \ref{tab:app_all_spectra}. The addition of the emission line significantly improved the quality of the fit. This fit to the spectrum is significantly preferred to a combination of plasma models and could indicate that the X-ray emission in this case is due to accretion onto a compact companion. An analysis of the changes in pulsations, flux, and spectrum of HD 162718 will be subject of a forthcoming paper by Webbe et al. (in prep).

\subsection{New cataclysmic variable candidates}
\label{subsec:res-newcv}
\begin{figure}
\centering
\includegraphics[width=\columnwidth]{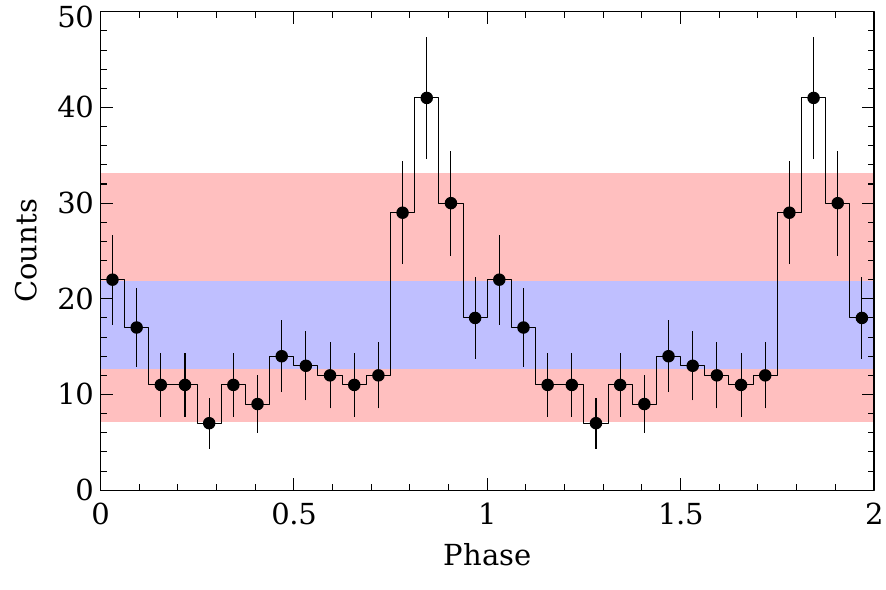}
\vspace{-0.25cm}
\caption{Pulsation profile for the CV candidate, source 1, 4XMM J173828.0-292842. X-ray events in the range 0.2--12.0\,keV are folded with a period of 12.98\,ks and bins of 810\,s. The blue and red bands indicate the 1$\sigma$ and 3$\sigma$ confidence intervals.}
\label{fig:src1-pulse}
\vspace{-0.25cm}
\end{figure}

Sources 1, 9, and 61 have been firmly classified through our analysis as CV candidates. Source 1, 4XMM J173828.0-292842, triggered an alert in observation 0886010101 as a first detection of an X-ray source. The source was located at RA=17h\ang{;38;27.90} DEC=-\ang{29;28;41.5}, with a position error of \ang{;;0.7}. A manual inspection identified a \emph{Gaia} counterpart 4060035990845753344, at a distance  of \ang{;;1.15} with a position error of \ang{;;0.01},  excluding the initial SIMBAD match against a long-period variable star, at a distance of $\sim$\ang{;;8.3} with a position error \ang{;;0.4} from the X-ray source location. This \emph{Gaia} source has no reliable parallax measurement and with $m_G = 20.78, $ it is only just detectable considering the nominal \emph{Gaia} detection limit of 20.7 \citep{hodgkin_gaia_2021}. The combination of optical colour ($BP-RP = 2.07$) and estimated X-ray to optical flux ratio $log(F_X/F_{Opt})\sim0.46$ identifies the source as a potential CV. The spectrum was well fit by a thermal bremsstrahlung model with a characteristic temperature $kT=2.82, $ with an uncertainty that cannot be constrained. This fit can be improved with the addition of an emission line at 6.7\,keV. All the spectral fits to the observation of 4XMM J173828.0-292842 are given in Table \ref{tab:app_all_spectra}. There were no clear features in the light curve in the full energy band, 0.2--12\,keV, a soft band (0.2--2\,keV), or a hard (2--12\,keV) band. The epoch-folding and $Z^2_n$ searches in the frequency range $3.2\times10^{-5}-5$\,Hz identified one 5$\sigma$-significant feature at $\sim7.70\times10^{-5}$\,Hz in an observation with exposure time 31.1\,ks. The X-ray events, folded at a period of 12.98\,ks ($f=7.70\times10^{-5}$\,Hz), display a distinct pulsation, as shown in Fig. \ref{fig:src1-pulse}.\\

\begin{figure}
\centering
\includegraphics[width=\columnwidth]{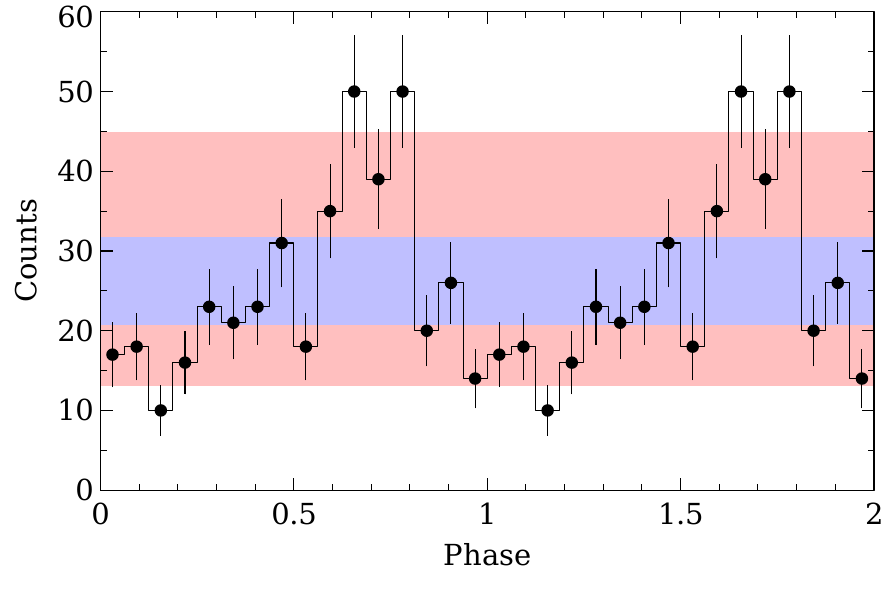}
\vspace{-0.25cm}
\caption{Pulsation profile for the CV candidate, source 9, 4XMM J173409.6-322946. X-ray events in the range 0.2--12.0\,keV are folded with a period of 8.21\,ks and bins of 256s. The blue and red bands indicate the 1$\sigma$ and 3$\sigma$ confidence intervals.}

\label{fig:src9-pulse}
\vspace{-0.25cm}
\end{figure}

Source 9, 4XMM J173409.6-322946, triggered an alert in observation 0886040201 as a first detection of an X-ray source. There was no identified SIMBAD association, but a potential optical counterpart, \emph{Gaia} source 4054630516785955200, was identified. The optical counterpart to 4XMM J173409.6-322946 is located at a distance of $\sim$640\,pc, which implies an X-ray luminosity of $\sim5.1\times10^{30}$erg\,s$^{-1}$. The combination of optical colour ($BP-RP = 2.96$) and  estimated X-ray to optical flux ratio of $log(F_X/F_{Opt})\sim0.01$ identified the source as a potential CV. The spectrum was well fitted with a thermal bremsstrahlung model with a characteristic temperature of $kT=4.61^{+30.56}_{-2.92}$. For all spectral fits to the observation of 4XMM J173409.6-322946 see Table \ref{tab:app_all_spectra}. There were no clear features in the light curve in the full energy band, 0.2--12\,keV, or a soft (0.2--2\,keV) or hard (2--12\,keV) band. Epoch folding and $Z^2_n$ searches in the frequency range $4.8\times10^{-5}-5$\,Hz identified one 5$\sigma$-significant feature at $\sim1.2\times10^{-4}$\,Hz in an observation of a duration of 20.8\,ks. The X-ray events, folded at a period of 8.21\,ks ($f=1.2\times10^{-4}$\,Hz), display a distinct pulsation, as shown in Fig. \ref{fig:src9-pulse}. \\

\begin{figure}
\centering
\includegraphics[width=\columnwidth]{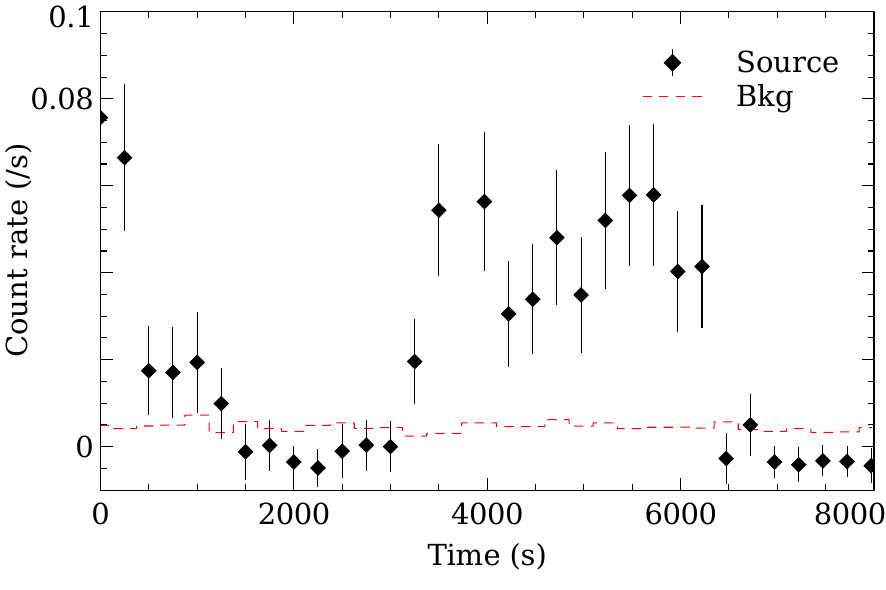}
\vspace{-0.25cm}
\caption{Light curve for the CV candidate, source 61, 2CXJ1800-2403. The light curve is for events in the energy range 0.2--12.0\,keV, with time bins of 250\,s. The source is shown by black diamonds, while the background is shown as a red dashed line.}
\label{fig:src61-lc}
\vspace{-0.25cm}
\end{figure}

Source 61, was identified as being in a high-flux state in observation 0932201101 and it was associated with the \emph{Chandra} source 2CXO J180012.0-240359 (henceforth, 2CXJ1800-2403). The X-ray light curve for the source during this observation showed clear variability, with two brighter periods being separated by a quiescent period of duration $\sim$3\,ks. The light curve for the observation is shown in Fig. \ref{fig:src61-lc}. The light curve appears to show one bright period, which may have begun before the start of the observation, and a second of duration $\sim$3\,ks, followed by a further quiescent period extending to the end of the observation. This manual inspection would suggest a period of approximately $P\sim5$\,ks. A search for optical and IR counterparts found no matches in the \emph{Gaia}, 2MASS, or AllWISE catalogues. The spectrum of 2CXJ1800-2403 was aptly fit by a thermal bremsstrahlung model with a characteristic temperature of $kT=7.08^{+16.57}_{-3.37}$ and an absorbing column density of $N_H=0.59^{+0.34}_{-0.21} \times10^{22}$\,cm$^{-2}$.  All the spectral fits to the observation of 2CXJ1800-2403 are given in Table \ref{tab:app_all_spectra}. The temperature is consistent with that expected for CVs, while the derived column density would imply a source distance on the order of $\sim$1.5\,kpc, assuming there is no local absorption, as per the nH3D tool\footnote{\url{http://astro.uni-tuebingen.de/nh3d/nhtool}} \citep{doroshenko_3d-n_rm_2024}. At this distance, the X-ray luminosity would be $log(L_X)=31.73$ and even when accounting for optical extinction along this line of sight of the order $A_V=1.7$, we can place a lower limit on the X-ray to optical flux ratio of $log(F_X/F_{Opt})\geq-0.28$. A search for periodicity was conducted using Fourier products, epoch folding and $Z^2_n$ searches, but no significant features were found. This is unsurprising given that the light curve appears to show that fewer than two possible cycles have been captured within the observation and, thus, no robust estimation of the periodicity is possible.

All three of these candidates had previous, constraining, non-detections during XMM observations, and source 61 was also previously detected at an even lower flux by \emph{Chandra}. Flux variations over more than one order of magnitude are possible for magnetic and non-magnetic CVs, as described in Sect. \ref{sec:intro}. The X-ray spectral characteristics are consistent with those of CVs \citep[e.g.][]{kuulkers_x-rays_2006,mukai_x-ray_2017}, while their optical and X-ray fluxes are within the ranges observed in similar X-ray CV samples \citep[e.g.][]{schwope_first_2024,galiullin_searching_2024}. The potential periodicities seen in sources 1 and 9 were only sampled over two full cycles and, thus, further monitoring would be required to confirm their existence; however, these results are consistent with the periods of other CVs identified in other Galactic samples \citep[e.g.][]{inight_catalogue_2023,galiullin_searching_2024}. As such, further monitoring of these sources would be needed to determine the strengths of the magnetic field in each case and the exact nature of the accreting white dwarf system.

\subsection{New X-ray binary candidate}
\label{subsec:res-newxrb}
\begin{figure}
\centering
\includegraphics[width=\columnwidth]{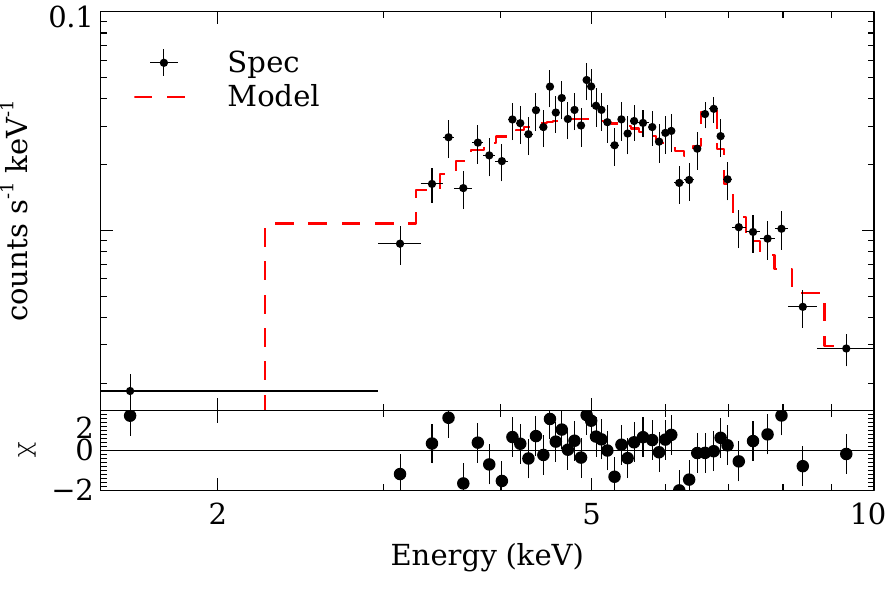}
\caption{Model fit to the spectrum of the XRB candidate, source 57, 4XMM J175033.4-264858. The spectrum is fit by an absorbed powerlaw model with an emission line fixed at 6.7\,keV. The observed spectrum is shown as black circles with associated error bars while the model is plotted as a red dashed line.}
\label{fig:src57-spec}
\vspace{-0.25cm}
\end{figure}

Source 57, 4XMM J175033.4-264858, was identified as being in a high-flux state in observation 0932190801. It was associated with the \emph{Chandra} source 2CXO J175033.4-264858 and with the SIMBAD IR source IRAS 17474-2648 by the STONKS pipeline. There is no visible optical counterpart for the source, which implies a lower limit on the X-ray to optical flux ratio of $log(F_X/F_{Opt})\geq0.93$, although this could be reduced due to significant extinction in the optical band. The X-ray light curve is dominated by harder energy photons, with energies greater than 2\,keV, which shows significant short term variability. A search for periodicity found no significant features in Fourier products or through epoch folding or $Z^2_n$ searches. The X-ray spectrum is best fitted with an absorbed non-thermal power law component with photon index $\alpha = 1.27^{+0.43}_{-0.39}$, column density of $N_H=18.42^{+3.82}_{-3.38} \times10^{22}$\,cm$^{-2}$, and an emission line centred at 6.7\,keV, which could correspond to an ionised Fe feature. Plasma models are strongly disfavoured in this case. All the spectral fits to the observation of 4XMM J175033.4-264858 are given in Table \ref{tab:app_all_spectra}. The fit to the spectrum is shown in Fig. \ref{fig:src57-spec}. 

This combination of the spectral fit, absence of optical counterpart, lack of pulsations, and X-ray-to-optical flux ratio indicate that source 57 could be a candidate XRB, likely in the low/hard state, if not in quiescence. There is significant optical extinction in the direction of the source, potentially of more than ten magnitudes making any determination as to a likely donor counterpart difficult, but does not preclude a high mass companion. The lack of pulsations and a higher luminosity potentially above 10$^{33}$\,erg\,s$^{-1}$ disfavour a CV interpretation and, in the absence of pulsations, there is no evidence to indicate whether the compact object is a neutron star or black hole. The very high column density, estimated to be $N_H \geq 10^{23}$\,cm$^{-2}$, could indicate the existence of a local absorber. This might indicate a high-mass XRB system being fed by winds from a giant companion, which temporarily increase absorption along the line of sight, although this would be equally consistent with a low-mass XRB. The high levels of optical extinction along the line of sight will hinder the possibility of determining the nature of this system by identification of the donor object. Instead, X-ray monitoring of 4XMM J175033.4-264858 is necessary to further understand the nature of this specific object by tracking the limits of the flux evolution and accompanying changes in the spectral shape and absorption.

\section{Conclusion}
\label{sec:conc}

In this analysis, we  present an application of the STONKS pipeline for the detection of highly variable X-ray sources to an \emph{XMM-Newton} survey of the Galactic plane. The pipeline was used to process source lists from 218 observations and produced 142 alerts for variable sources. Screening for observational artefacts resulted in the identification of 78 true alerts for variability associated with 70 sources. These sources were classified through considering a combination of X-ray variability within observations, X-ray spectral fitting, previous X-ray source detections, as well as optical and NIR-MIR detections. This approach enabled these detections, with a firm classification achieved for 38 of the 70 sources. 

In particular we identified one new magnetar candidate, detected at the onset of an outburst, 4XJ1751-2759, along with evolving pulsations in the $\gamma$-Cas analogue HD 162718, three new CV candidates, with periods of 13.0\,ks, 8.2
    \,ks and $\sim$5\,ks, and one new XRB candidate. We  also identified a population of CV candidates located at greater distances than those routinely selected in optical surveys. Extending the population of CVs to greater distances will help to further constrain stellar evolution pathways. The application of STONKS to this survey has demonstrated the potential of this pipeline to identify highly variable X-ray sources. With the implementation of STONKS as part of the XMM processing pipeline in proposal cycle AO24, new alerts are being made available to the community as soon as they become available after processing through the \textit{XMM-Newton} Survey Science Centre\footnote{\url{http://flix.irap.omp.eu/stonks}}. 
\\

\section*{Data availability}\par
Tables C1, C2, C3, and D1 are only available in electronic form at the CDS via anonymous ftp to cdsarc.u-strasbg.fr (130.79.128.5) or via http://cdsweb.u-strasbg.fr/cgi-bin/qcat?J/A+A/

\begin{acknowledgements}
Authors RW and NW acknowledge support from the CNES for this work. This project has received funding from the European Union’s Horizon 2020
research and innovation programme under grant agreement n°101004168, the
XMM2ATHENA project. GP acknowledges financial support from the European Research Council (ERC) under the European Union's Horizon 2020 research and innovation program HotMilk (grant agreement No. 865637), support from Bando per il Finanziamento della Ricerca Fondamentale 2022 dell'Istituto Nazionale di Astrofisica (INAF): GO Large program and from the Framework per l'Attrazione e il Rafforzamento delle Eccellenze (FARE) per la ricerca in Italia (R20L5S39T9). SM acknowledges support by the National Science Foundation Graduate Research Fellowship under Grant No. DGE 2036197 and the Columbia University Provost Fellows Program. SM acknowledges support from grant NASA ADAP 80NSSC24K0666. This research has made use of data obtained from the Chandra Source Catalog, provided by the Chandra X-ray Center (CXC). This research has made use of data obtained from the 4XMM \textit{XMM-Newton} serendipitous source catalogue compiled by the \textit{XMM-Newton} Survey Science Centre consortium. This work has made use of data from the European Space Agency (ESA) mission {\it Gaia} (\url{https://www.cosmos.esa.int/gaia}), processed by the {\it Gaia} Data Processing and Analysis Consortium (DPAC, \url{https://www.cosmos.esa.int/web/gaia/dpac/consortium}). Funding for the DPAC has been provided by national institutions, in particular the institutions participating in the {\it Gaia} Multilateral Agreement. This publication makes use of data products from the Two Micron All Sky Survey, which is a joint project of the University of Massachusetts and the Infrared Processing and Analysis Center/California Institute of Technology, funded by the National Aeronautics and Space Administration and the National Science Foundation. This publication makes use of data products from the Wide-field Infrared Survey Explorer, which is a joint project of the University of California, Los Angeles, and the Jet Propulsion Laboratory/California Institute of Technology, funded by the National Aeronautics and Space Administration.

\end{acknowledgements}

\bibliographystyle{aa}

\bibliography{STONKS_GalPlane}

\begin{appendix}
\section{Source classification methodology}
\label{app:class-criteria}
We present here the full details of the classification methodology related to specific source classes. Due to the sparseness of multi-wavelength information for many of our sources and with the extinction across all bands expected to be very high as the survey is covering the Galactic plane, we considered these criteria in a non-discriminatory manner.

\begin{itemize}
    \item Star:
    \begin{itemize}
        \item Source is classified as a star in the \textit{Gaia} DR3 catalogue.
        \item Fit the multi-wavelength information to known stellar templates using the \texttt{ARIADNE} module for \texttt{python}, and compare with Gaia detections.
        \item Confirm that stellar parameters from \texttt{ARIADNE} are consistent with the Gaia source location distance and expected extinction.
        \item Verify that X-ray flux and luminosity are such that $log(F_X/F_{Opt}) \lesssim -2$, $log(L_X) \lesssim 32$ \citep{pye_survey_2015,schmitt_forbidden_2024}, and that $log(L_x/L_{Bol}) \lesssim -3$ \citep{pizzolato_stellar_2003}.
        \item Verify that the absorbed \texttt{apec} model parameters have physically meaningful and reasonable values.
    \end{itemize}
    \item YSO:
    \begin{itemize}
        \item X-ray source location is within that of known star-forming regions \citep{avedisova_catalog_2002}.
        \item Identification of the source as a YSO by MYStIX \citep{broos_identifying_2013}.
        \item Identify an excess in the NIR spectrum of the source, using the criteria of \citet{zeidler_vista_2016}. ($m_J - m_H$) > 0.05, ($m_H - m_{K_S}$) > 0.05 and ($m_J - m_H$) < 1.86 $\times$ ($m_H - m_{K_S}$) - 0.05.
        \item Luminosity at the estimated source distance is such that $log(L_X) \lesssim 33$ and that $log(L_x/L_{Bol}) \lesssim -1$ \citep{barrado_xmm-newton_2011,getman_x-ray_2021}
        \item Fit the X-ray spectrum to that of an absorbed \texttt{apec} plasma \citep{getman_chandra_2006}, or \texttt{raymond} \citep{winston_chandra_2018} hot gas model, with abundances set to 0.3, and confirm that the spectrum has physically meaningful and reasonable parameters for a YSO.
    \end{itemize}
    \item CV
    \begin{itemize}
        \item Identify periodicity in the X-ray light curves within expected ranges for CVs, with orbital periods $\sim1.25-8$\,hours \citep{kuulkers_x-rays_2006,mukai_x-ray_2017}.
        \item Verify that $log(F_X/F_{Opt}) \gtrsim 1 - 3.5(BP-RP)$ using the methodology of \citet{galiullin_searching_2024} and \citet{rodriguez_active_2024}.
        \item Fit the X-ray spectrum to that of a thermal bremsstrahlung \texttt{bremss} emission model \citep{kuulkers_x-rays_2006} with physically meaningful and reasonable parameters ($kT\sim 1-20$\,keV; e.g. \citealt{kuulkers_x-rays_2006,mukai_x-ray_2017}).
        \item Source is classified as a white dwarf in Gaia DR3 catalogue.
    \end{itemize}
    \item XRB
    \begin{itemize}
        \item Verify that $log(F_X/F_{Opt}) \gtrsim 1$ using an upper limit on the source optical magnitude of 20.7, where necessary, as the Gaia detection limit \citep{hodgkin_gaia_2021}.
        \item Verify that $log(L_X) \gtrsim 31$ \citep[e.g.][]{remillard_x-ray_2006,bahramian_low-mass_2023,fornasini_high-mass_2023} using the distance estimate of 8\,kpc if no other is available.
        \item Fit the spectrum to a combined absorbed blackbody and powerlaw model (\texttt{bbody + powerlaw}) with expected normalisations and parameter values to match those expected in XRBs in hard or soft states \citep{remillard_x-ray_2006}.
    \end{itemize}
\end{itemize}

\section{Luminosity distribution for firmly classified STONKS alert sources}

\begin{figure}[h]
\centering
\includegraphics[width=\columnwidth]{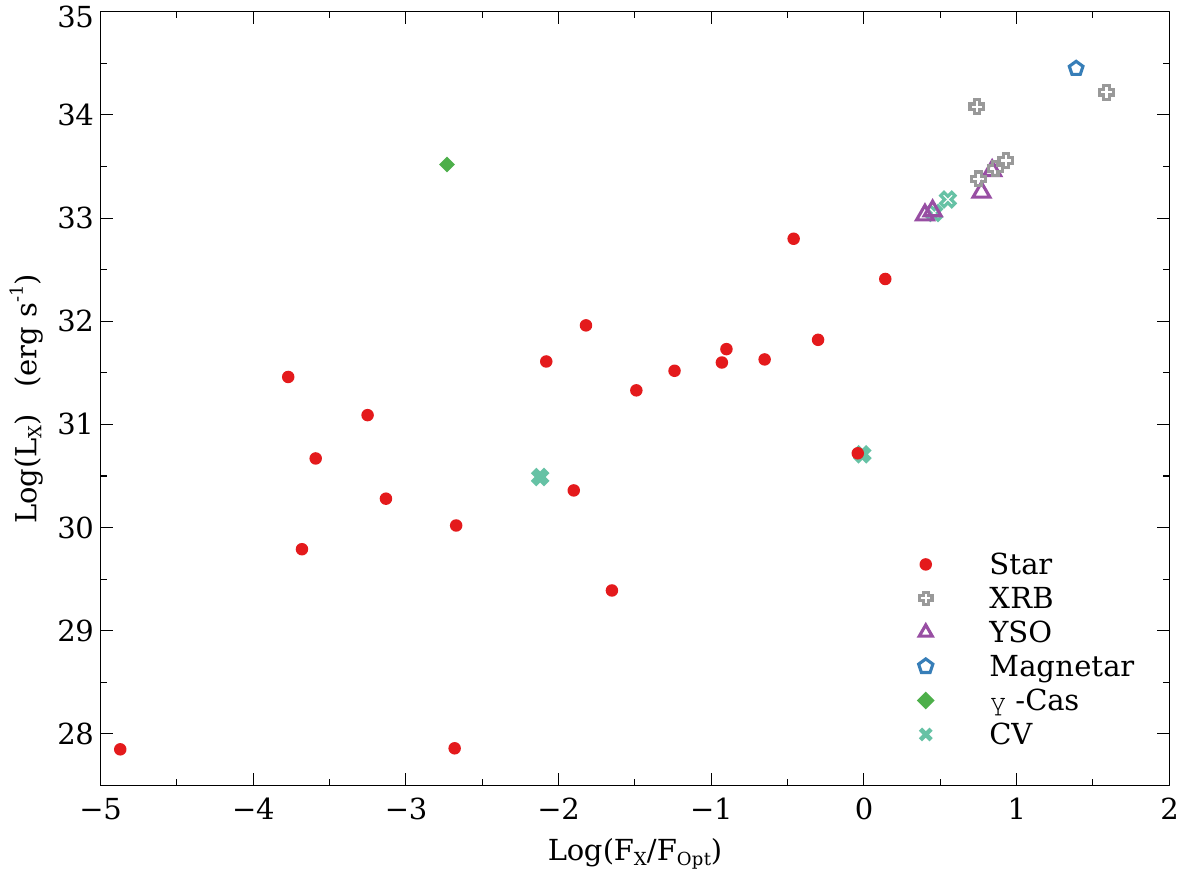}
\caption{Optical to X-ray flux ratios and X-ray luminosities for the 38 sources with a firm classification as per our methodology outlined in Sect. \ref{subsec:methods-class}. We plot the different classes as: star - red circle; XRB - grey cross; YSO - purple triangle; CV - cyan cross; magnetar candidate - blue pentagon; $\gamma$-Cas star - green diamond. Note that the apparent linear trend for some YSO, CV, and XRB sources is due to the lack of a MW counterpart with an associated distance estimate, or the lack of any MW counterpart. In these cases we use a distance of 8\,kpc in deriving a luminosity estimate and the \emph{Gaia} detection limit as an upper limit for the optical flux. Sources without a distance estimate from a MW counterpart are shown as un-filled symbols.}
\label{fig:firm-class-fxopt-lx}
\end{figure}

\section{True alert details}
Here, we present the first ten lines of three tables, each of which is available in full through Vizier. These three tables list the details of the STONKS alert, multi-wavelength associations, and spectral fitting to X-ray observations respectively.

\begin{sidewaystable}
\centering
\caption{X-ray properties of STONKS alert sources from Multi-Year Heritage Project observations of the Galactic Plane. Locations and flux values are as per the source details in the XMM Science Archive.}
\label{tab:app_screened_alerts_xray1}
\begin{tabular}{lcccccccccc}
\hline\hline
Alert Src. No. & OBSID & Source & RA & DEC & Error (") & 4XMM DR14 Ident. & Alert Class & Spectral & Flux & True Alert \\
\hline
1 & 0886010101 & 12 & 17:38:27.90 & -29:28:41.5 & 0.7 & J173828.0-292842 & FD & Y & 2.58 & Y \\
2 & 0886010401 & 3 & 17:45:24.98 & -30:47:50.7 & 0.4 & J174525.0-304750 & FD & Y & 2.46 & Y \\
3 & 0886010401 & 4 & 17:45:55.32 & -30:42:40.3 & 0.6 & J174555.4-304240 & FD & N & 0.90 & Y \\
4 & 0886010401 & 9 & 17:46:23.34 & -20:37:58.6 & 0.8 & J174623.4-303758 & H & Y & 1.60 & N \\
5 & 0886010501 & 3 & 17:44:05.03 & -30:39:54.2 & 0.5 & J174405.0-303952 & FD & Y & 4.57 & Y \\
\multirow{2}{6em}{6} & 0886010801 & 11 & 17:38:41.46 & -30:14:57.2 & 0.9 & \multirow{2}{*}{J173841.4-301458} & H & Y & 2.08 & Y \\
 & 0886121201 & 10 & 17:38:41.48 & -30:14:59.7 & 0.9 &  & H & Y & 2.35 & Y \\
7 & 0886011001 & 10 & 17:39:59.62 & -30:38:46.8 & 1.2 & J173959.5-303847 & L & N & 0.28 & N \\
8 & 0886020501 & 23 & 17:39:47.40 & -31:01:51.3 & 1.3 & J173947.4-310152 & FD & Y & 4.12 & Y \\
9 & 0886040201 & 1 & 17:34:09.69 & -32:29:47.2 & 0.6 & J173409.6-322946
 & FD & N & 1.85 & Y \\
\multirow{3}{6em}{10} & 0886040201 & 20 & 17:34:06.23 & -32:28:58.4 & 0.9 & \multirow{3}{*}{J173406.1-322857} & FD & Y & 1.54 & Y \\
 & 0934200401 & 3 & 17:34:06.24 & -32:28:53.1 & 1.2 &  & FD & Y & 5.02 & Y \\
 & 0934201001 & 19 & 17:34:06.22 & -32:28:53.4 & 2.3 &  & FD & Y & 3.03 & Y \\
\hline
\end{tabular}
\tablefoot{Alert classes align with those possible through the STONKS pipeline: H - High flux state, L - Low flux state, FD - First detection, PV - Past variability. The spectral column identifies those sources where the spectrum deviates significantly from the assumptions made in pipeline flux estimations. Fluxes are stated in units of $10^{-13}$\,erg\,s$^{-1}$\,cm$^{-2}$. The final column identifies those STONKS alerts which may be as the result of spectral assumptions between different instruments and not truly caused by significant flux variability: Y - Likely a true flux variation, N - Likely a false variation due to spectral assumptions, ? - Possibly a false variation. The full Table is available at the CDS.}
\end{sidewaystable}

\begin{sidewaystable}
\centering
\caption{Multi-wavelength properties of counterparts to STONKS alert sources from Multi-Year Heritage Project observations of the Galactic Plane.}
\label{tab:app_screened_alerts_mw1}
\begin{tabular}{lcccccccccc}
\hline\hline
Number & SIMBAD & Gaia & $m_G$ & Sep (") & 2MASS & $m_J$ & Sep (") & AllWISE & $m_{W1}$ & Sep (") \\
\hline
1 & 2MASS J17382849-2928445 & 4060035990845753344 & 20.8 & 1.2 & -- & -- & -- & -- & -- & -- \\
2 & TYC 7377-839-1 & -- & -- & -- & -- & -- & -- & -- & -- & -- \\
3 & -- & -- & -- & -- & -- & -- & -- & -- & -- & -- \\
4 & [JBN2011] 1005 & -- & -- & -- & -- & -- & -- & -- & -- & -- \\
5 & -- & -- & -- & -- & 17440506-3039522 & 14.0 & 2.1 & J174405.05-303952.1 & 10.0 & 2.2 \\
6 & -- & -- & -- & -- & -- & -- & -- & -- & -- & -- \\
7 & HD 316176 & 4055398766191461376 & 9.6 & 3.4 & 17395979-3038469 & 8.2 & 2.6 & J173959.76-303848.6 & 7.3 & 2.8 \\
8 & -- & -- & -- & -- & 17394755-3101532 & 15.4 & 2.8 & -- & -- & -- \\
9 & -- & 4054630516785955200 & 20.0 & 2.0 & -- & -- & -- & -- & -- & -- \\
10 & -- & 4054631272700204672 & 18.4 & 4.2 & -- & -- & -- & -- & -- & -- \\
\hline
\end{tabular}
\tablefoot{The second column states those counterparts identified in SIMBAD by the STONKS pipeline, and should be treated with appropriate caution. The remaining nine columns contain source IDs and magnitudes for the most likely counterparts as identified by individual visual inspection. Magnitudes are stated as apparent, rather than absolute, due to the lack of completeness of distance estimates. The full Table is available at the CDS.}
\end{sidewaystable}

\begin{sidewaystable}
\centering
\caption{Classifications of X-ray selected STONKS alert sources. We list here the STONKS alert  class and available multi-wavelength counterparts to the X-ray sources.}
\label{tab:app_all_class1}
\begin{tabular}{cccccccccccccccc}
\hline\hline
No. & Alert & MW (O/N/M) & $d$ (kpc) & $log(L_X)$ & $log(F_X/F_{\text{Opt}})$ & Short Variable & Spectrum & $kT$ & $kT_2$ & $kT_3$ & Abund & $\alpha$ & Src Class & Quality \\
\hline
1 & FD & O & -- & 33.05 & 0.46 & Y & \texttt{bremss} & 1.22 & -- & -- & -- & -- & CV & F \\
2 & FD & -- & -- & 32.97 & $\geq$0.35 & N & \texttt{bbody} & 1.66 & -- & -- & -- & -- & XRB & ? \\
3 & FD & -- & -- & 32.80 & $\geq$0.18 & N & \texttt{ap} & 43.25 & -- & -- & 1.0 & -- & Star & ? \\
4 & H & -- & -- & 32.95 & $\geq$0.33 & N & \texttt{powerlaw} & -- & -- & -- & -- & 1.38 & XRB & ? \\
5 & FD & N/M & -- & 33.64 & $\geq$1.01 & N & \texttt{bb + pl} & 0.22 & -- & -- & -- & 1.07 & XRB & ? \\
\multirow{2}{*}{6} & H & \multirow{2}{*}{--} & \multirow{2}{*}{--} & 33.06 & $\geq$0.43 & N & \texttt{bbody} & 2.27 & -- & -- & -- & -- & \multirow{2}{*}{XRB} & \multirow{2}{*}{?} \\
 & H &  &  & 33.43 & $\geq$0.81 & N & \texttt{bbody} & 1.12 & -- & -- & -- & -- &  &  \\
7 & L & O/N/M & 0.05 & 27.85 & -4.87 & Y & \texttt{apec} & 0.27 & -- & -- & 1.0 & -- & Star & F \\
8 & FD & N & -- & 33.32 & $\geq$0.69 & N & \texttt{powerlaw} & -- & -- & -- & -- & 0.31 & XRB & ? \\
9 & FD & O & 0.64 & 30.71 & -0.01 & Y & \texttt{bremss} & 4.62 & -- & -- & -- & -- & CV & F \\
\multirow{3}{*}{10} & FD & \multirow{3}{*}{O} & \multirow{3}{*}{0.97} & 31.10 & -0.63 & N & \texttt{ap + ap} & 0.11 & 3.35 & -- & 1.0 & -- & \multirow{3}{*}{Star} & \multirow{3}{*}{?} \\
& FD &  &  & 31.51 & -0.22 & N & \texttt{apec} & 16.48 & -- & -- & 1.0 & -- &  &  \\
& FD &  &  & 31.33 & -0.40 & N & \texttt{apec} & 0.34 & -- & -- & 1.0 & -- &  &  \\
\hline
\end{tabular}
\tablefoot{Distance measures ($d$) are reported in units of kpc, and the luminosity estimates ($log(L_X)$) are calculated using these distances in units of erg\,s$^{-1}$. Where no distance estimate is provided, luminosities are estimated using a distance of 8\,kpc, the approximate distance to the Galactic centre. The column $log(F_X/F_{\text{Opt}})$ reports the X-ray-to-optical flux ratio, or the lower limit of such if there is no optical counterpart with a detection limit of $m_G$ 20.7. The Short Variable column indicates those sources where significant variability was detected within the light curve of the source for which the alert was created. The Spectrum column indicates the model which best characterises the source in the context of its classification. The $kT$ (also $kT_2$ and $kT_3$), Abundance, and $\alpha$ columns indicate the characteristic temperature(s), solar abundance as fixed for model fitting, or power law index derived from the spectral fit as required. The final two columns indicate the likely source class, and the strength of this classification (F - Firm, ? - Provisional). The full Table is available at the CDS.}
\end{sidewaystable}

\section{Spectral fitting}

Here, we present the first three lines of a table, which is available in full through Vizier. This table lists the details of the spectral fitting to X-ray observations of the STONKS-selected X-ray sources.

\renewcommand{\arraystretch}{1.2}
\begin{sidewaystable}
\centering
\caption{Results of spectral fitting to X-ray selected STONKS alert sources. We list here the STONKS alert class and parameters derived from fitting the X-ray spectra to X-ray sources.}
\label{tab:app_all_spectra}
\begin{tabular}{ccc|ccccccc|cccc}
\hline\hline
No. & Alert & OBSID & Model & $N_{\text{H}}$ & $kT$ & $kT_2$ & $kT_3$ & Abund & $\alpha$ & dof & $\chi^2_{\nu}$ & $N$ & $C$-stat\\
\hline
\multirow{5}{*}{1} & \multirow{5}{*}{FD} & \multirow{5}{*}{0886010101} & \texttt{bbody} & $68.38^{+217.29}_{-\infty}$ & $1.22^{+6.99}_{-0.93}$ & -- & -- & -- & -- & 7 & 1.29 & -- & -- \\
 &  &  & \texttt{powerlaw} & $84.38^{+119.10}_{-\infty}$ & -- & -- & -- & -- & $3.86^{+\infty}_{-6.46}$ & 7 & 1.30 & -- & -- \\
 &  &  & \texttt{apec} & $39.50^{+34.30}_{-20.35}$ & $57.59^{+\infty}_{-51.86}$ & -- & -- & 1.0 & -- & 7 & 1.40 & -- & -- \\
 &  &  & \texttt{bremss} & $78.16^{+223.95}_{-56.16}$ & $2.81^{+\infty}_{-\infty}$ & -- & -- & -- & -- & 7 & 1.30 & -- & -- \\
 &  &  & \texttt{bremss + gauss} & $0.77^{+8.27}_{-0.77}$ & $22.79^{+\infty}_{-22.73}$ & 6.7 & -- & -- & -- & 5 & 0.86 & -- & -- \\
\hline
\multirow{4}{*}{2} & \multirow{4}{*}{FD} & \multirow{4}{*}{0886010401} & \texttt{bbody} & $1.19^{+2.39}_{-1.03}$ & $1.66^{+1.03}_{-0.58}$ & -- & -- & -- & -- & 10 & 0.76 & -- & -- \\
 &  &  & \texttt{powerlaw} & $2.75^{+4.67}_{-2.10}$ & -- & -- & -- & -- & $1.08^{+1.70}_{-1.11}$ & 10 & 0.72 & -- & -- \\
 &  &  & \texttt{apec} & $4.12^{+3.53}_{-1.31}$ & $7.80^{+\infty}_{-5.58}$ & -- & -- & 1.0 & -- & 10 & 0.51 & -- & -- \\
 &  &  & \texttt{bremss} & $3.05^{+1.42}_{-0.96}$ & $199.36^{+\infty}_{-\infty}$ & -- & -- & -- & -- & 10 & 0.72 & -- & -- \\
\hline
\multirow{4}{*}{3} & \multirow{4}{*}{FD} & \multirow{4}{*}{0886010401} & \texttt{bbody} & $1.40^{+2.80}_{-\infty}$ & $1.70^{+2.31}_{-0.72}$ & -- & -- & -- & -- & 9 & 1.80 & -- & -- \\
 &  &  & \texttt{powerlaw} & $3.14^{+3.83}_{-2.26}$ & -- & -- & -- & -- & $1.38^{+1.80}_{-1.33}$ & 9 & 1.62 & -- & -- \\
 &  &  & \texttt{apec} & $2.98^{+2.57}_{-1.17}$ & $63.93^{+\infty}_{-\infty}$ & -- & -- & 1.0 & -- & 9 & 1.63 & -- & -- \\
 &  &  & \texttt{bremss} & $2.95^{+2.92}_{-1.22}$ & $95.24^{+\infty}_{-\infty}$ & -- & -- & -- & -- & 9 & 1.63 & -- & -- \\
\hline
\end{tabular}
\tablefoot{The Model column indicates the model which has been combined with ISM absorption (\texttt{tbabs*}) to provide the spectral fit. The $N_{\text{H}}$ column gives the neutral hydrogen column density in units of $\times10^22$\,cm$^{-2}$. The $kT$ (also $kT_2$ and $kT_3$), Abundance, and $\alpha$ columns indicate the characteristic temperature(s) in units of keV, solar abundance as fixed for model fitting, or power law index derived from the spectral fit as required. The final four columns indicate the quality of the fit. They either give the give the number of degrees of freedom and the reduced chi-squared value, or the number of photons and the Cash stat value. Where the errors on physical parameters cannot be bound by fitting this is indicated as $\infty$. The full Table is available at the CDS.}
\end{sidewaystable}

\end{appendix}

\end{document}